\documentclass[letterpaper]{article} 
\usepackage[draft]{aaai25}  
\usepackage{times}  
\usepackage{helvet}  
\usepackage{courier}  
\usepackage[hyphens]{url}  
\usepackage{graphicx} 
\usepackage{natbib}  
\usepackage{caption} 
\usepackage{algorithm}
\usepackage{algorithmic}
\usepackage{amsmath}
\usepackage{amssymb}
\usepackage{multirow}

\usepackage{newfloat}
\usepackage{listings}
\DeclareCaptionStyle{ruled}{labelfont=normalfont,labelsep=colon,strut=off} 
\floatstyle{ruled}
\newfloat{listing}{tb}{lst}{}
\floatname{listing}{Listing}
\usepackage{xspace}
\usepackage[table]{xcolor}
\usepackage{booktabs} 

\usepackage{subfigure}

\definecolor{successGreen}{rgb}{0.835,0.969,0.812}
\newcommand{\greencell}[1]{\cellcolor[rgb]{0.835,0.969,0.812}#1}

\def\eg{\emph{e.g.}\xspace} 

\def\ie{\emph{i.e.}\xspace}

\def\etc{\emph{etc}\xspace}

\newcommand{\nameFramework}{OTBR\xspace}

\newcommand{\vect}[1]{\boldsymbol{#1}}

\definecolor{darkred}{rgb}{0.7,0,0}
\definecolor{darkgreen}{rgb}{0,0.46,0}
\definecolor{purple}{rgb}{0.6,0,0.5}
\definecolor{cholocate}{HTML}{d2691e}
\definecolor{slateblue}{HTML}{6a5acd}

\newcommand{\fin}{\color{black}}

\newcommand{\eqdef}{\stackrel{\sf def}{=}}

\title{Fusing Pruned and Backdoored Models: Optimal Transport-based \\ Data-free Backdoor Mitigation}
\author{
    Weilin Lin\textsuperscript{\rm 1}, Li Liu\textsuperscript{\rm 1}\thanks{Corresponds to Li Liu (avrillliu@hkust-gz.edu.cn)}, Jianze Li\textsuperscript{\rm 2,3}, Hui Xiong\textsuperscript{\rm 1}
}
\affiliations{
    \textsuperscript{\rm 1}The Hong Kong University of Science and Technology (Guangzhou)\\
    \textsuperscript{\rm 2}Shenzhen Research Institute of Big Data\\
    \textsuperscript{\rm 3}The Chinese University of Hong Kong, Shenzhen

}

\usepackage{bibentry}

\begin{document}

\maketitle

\begin{abstract}

Backdoor attacks present a serious security threat to deep neuron networks (DNNs). Although numerous effective defense techniques have been proposed in recent years, they inevitably rely on the availability of either clean or poisoned data. In contrast, \textbf{data-free} defense techniques have evolved slowly and still lag significantly in performance. To address this issue, different from the traditional approach of pruning followed by fine-tuning, we propose a novel \textbf{data-free} defense method named \textit{\textbf{Optimal Transport-based Backdoor Repairing}} (\textit{\textbf{\nameFramework}}) in this work. 
This method, based on our findings on \textit{neuron weight changes (NWCs)} of random unlearning, uses \textit{optimal transport (OT)}-based model fusion to combine the advantages of both pruned and backdoored models. 
Specifically, we first demonstrate our findings that the NWCs of random unlearning are positively correlated with those of poison unlearning. Based on this observation, we propose a \textit{random-unlearning NWC pruning} technique to eliminate the backdoor effect and obtain a backdoor-free pruned model. Then, motivated by the OT-based model fusion, we propose the \textit{pruned-to-backdoored OT-based fusion} technique, which fuses pruned and backdoored models to combine the advantages of both, resulting in a model that demonstrates high clean accuracy and a low attack success rate. 
To our knowledge, this is the first work to introduce OT and model fusion techniques to the backdoor defense. 
Extensive experiments show that our method successfully defends against all seven backdoor attacks across three benchmark datasets, outperforming both state-of-the-art (SOTA) data-free and data-dependent methods. 
\end{abstract}

%

\section{Introduction}
\label{sec:intro}

Over the past decade, deep neural networks (DNNs) have become a crucial technology in various applications, including image recognition~\cite{parmar2014face, he2016deep}, speech processing~\cite{gaikwad2010review, maas2017building}, and natural language processing~\cite{chowdhary2020natural}, \etc. However, as the deployment of DNNs in sensitive and critical domains becomes more widespread, concerns regarding their security cannot be ignored. 
Among the numerous threats to DNNs, \textit{backdoor attacks} ~\cite{gu2019badnets, li2021invisible,wu2023adversarial} are particularly concerning. 
In these attacks, the attackers manipulate a small portion of the training data to implant a stealthy backdoor into a DNN, resulting in a \textit{backdoored model.} During inference, the backdoored model behaves anomalously when the input contains a pre-defined trigger pattern; otherwise, it performs normally.
This phenomenon is termed the \textit{backdoor effect}.
Such attacks may pose hidden security issues to real-world applications, such as unauthorized access to a system when a company develops its software using a third-party pre-trained model. 

\begin{figure}[t]
\centering
\includegraphics[width=0.48\columnwidth]{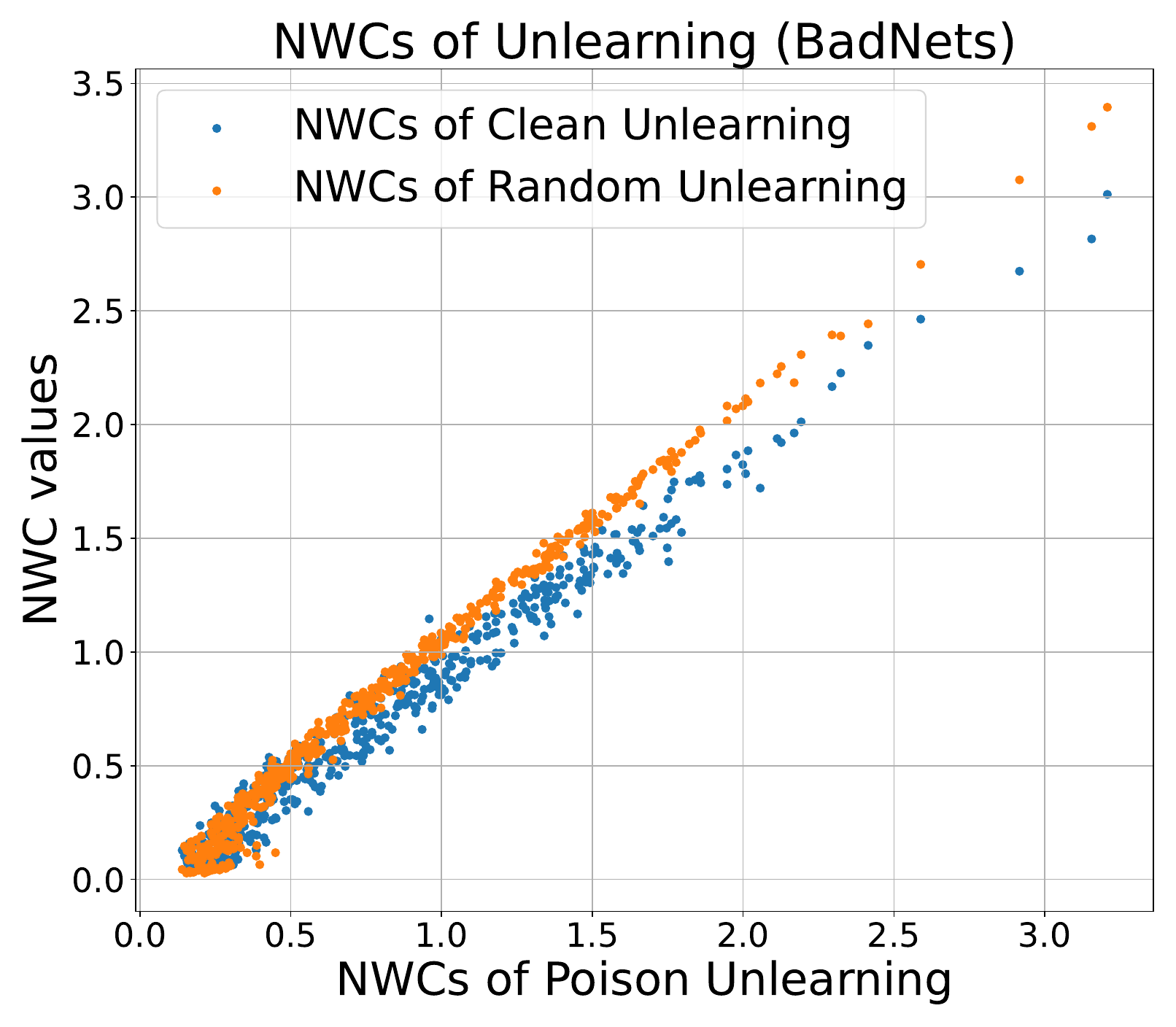} 
\includegraphics[width=0.48\columnwidth]{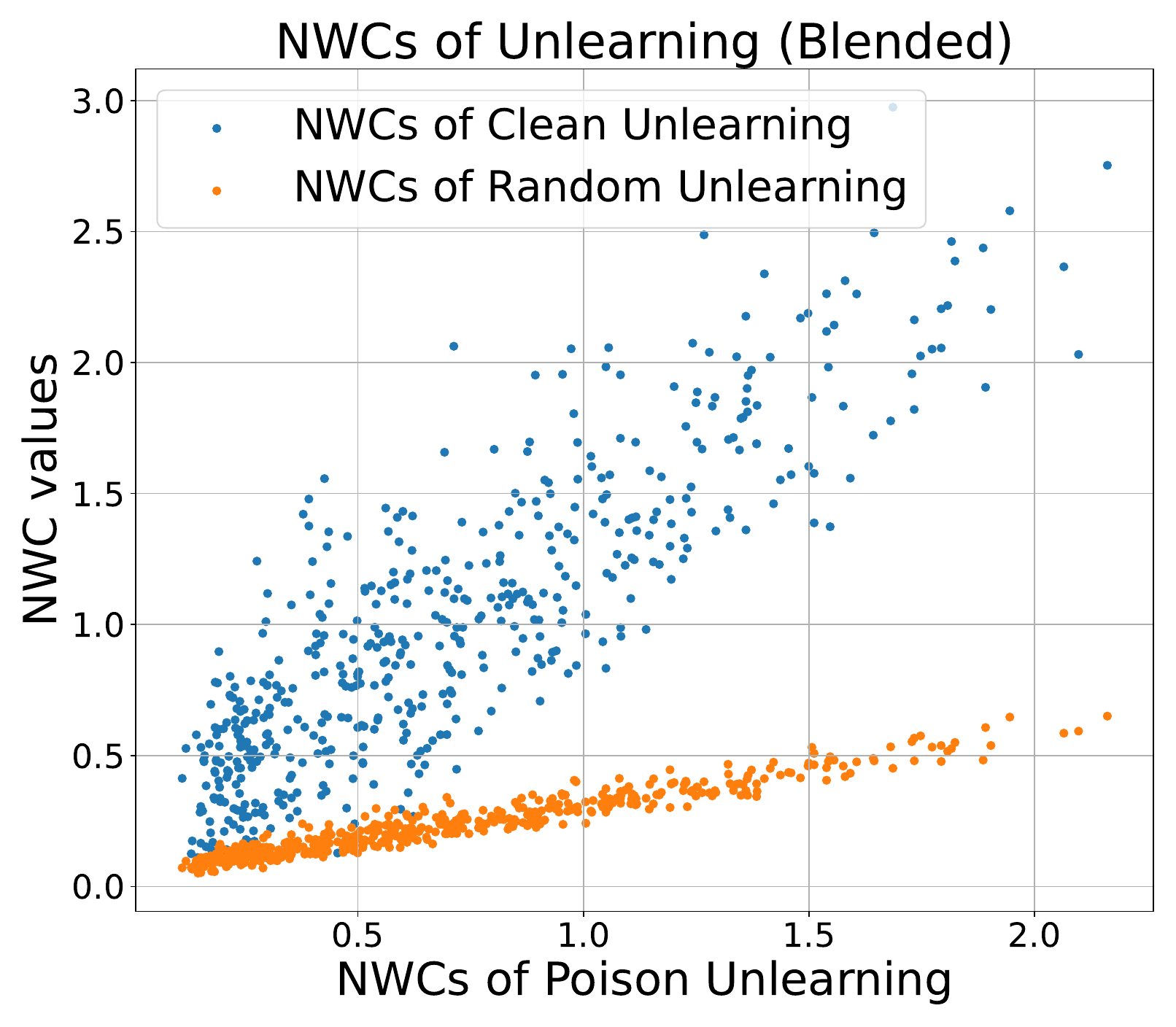}
\caption{Illustration of unlearning NWCs on BadNets~\cite{gu2019badnets} and Blended~\cite{chen2017targeted} attacks.  
The NWCs of both clean and random unlearning show a positive correlation with poison unlearning. The last convolutional layer is chosen for this illustration.   }
\label{fig:nwc}
\end{figure}

\begin{figure}[t]
\centering
\includegraphics[width=1\columnwidth]{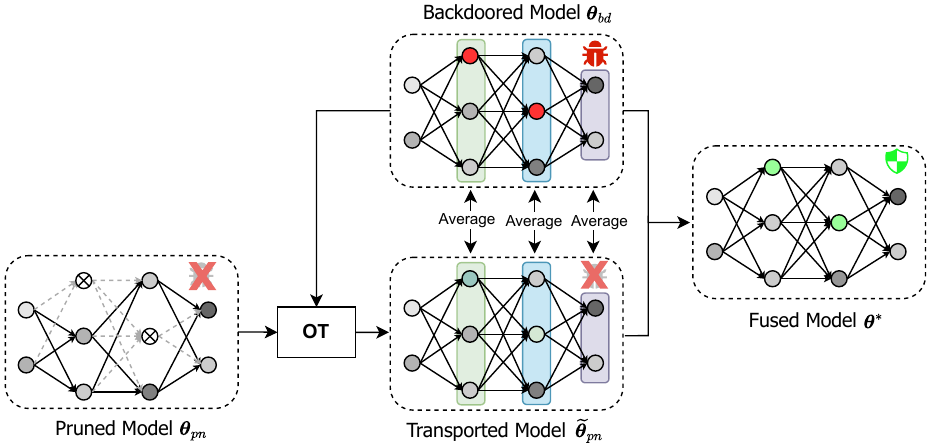} 
\caption{OT-based model fusion for backdoor defense. The pruned model is aligned with the backdoored model layer-by-layer using OT. Then the models are fused through a weighted averaging operation.}
\label{fig:overview}
\end{figure}

In recent years, as backdoor attack methods have evolved, \textit{backdoor defense} techniques have also seen significant growth. 
Various important techniques have been developed for backdoor defense, including pruning~\cite{wu2021adversarial}, unlearning~\cite{zeng2021adversarial}, and fine-tuning~\cite{zhu2023enhancing}, \etc.  However, most of these techniques rely on the availability of clean or poisoned data, which restricts their applicability to the aforementioned scenarios. 
Recent insights reveal a promising direction ~\cite{lin2024unveiling}: using \textbf{\textit{neuron weight changes (NWCs)}} of \textit{clean unlearning}\footnote{Unlearning the backdoored model on clean data.} to categorize the neurons into backdoor-related ones and clean ones\footnote{The larger NWC of a neuron, the more backdoor-related it is.}, based on an observation that the NWCs of unlearning clean and poisoned data are positively correlated. 
In this work, our extended findings reveal that using random noise for unlearning, termed as \textit{random unlearning}, brings a new similar insight: \textbf{the NWCs of random unlearning exhibit a positive correlation with those of poison unlearning} (as shown in Figure~\ref{fig:nwc}). This motivates us to adopt NWCs for \textbf{data-free} backdoor mitigation using only the generated random noise. 
Normally, after identifying backdoor-related neurons, pruning (or zero reinitialization) followed by fine-tuning is employed to eliminate the backdoor effect and restore the lost performance~\cite{liu2018fine}. 
However, this is infeasible in data-free scenarios since the subsequent fine-tuning requires clean data. If we only perform pruning using NWCs and simply skip the fine-tuning, the \textit{clean accuracy} (ACC) is prone to decrease by more than 10\%~\cite{lin2024unveiling}. Therefore, after pruning, it is necessary to develop a new data-free technique for performance recovery. 

Recently, \textit{model fusion}~\cite{li2023deep} has received increasing attention. 
It combines the weights of multiple models to integrate their capabilities into a single network. 
As one of the most representative works, OTFusion~\cite{singh2020model} employs \textit{optimal transport (OT)} to align model weights layer-by-layer before fusing two models through averaging. 
Following it, Intra-Fusion~\cite{theus2024towards} employs OT to integrate the functionality of pruned neurons with the remaining ones, aiming to maintain great performance after pruning. It can be seen that the aforementioned methods both demonstrate OT's inherent ability to preserve critical information during the fusion process.

Motivated by the above advancements in model fusion, in this work, we explore its potential to combine the high ACC of the backdoored model with the low \textit{attack success rate} (ASR) of the pruned model in a data-free manner.
Building on the foundation of NWC pruning and OT-based model fusion, we propose a novel data-free defense strategy called \textit{\textbf{O}ptimal \textbf{T}ransport-based \textbf{B}ackdoor \textbf{R}epairing} (\nameFramework), which fuses pruned and backdoored models. \nameFramework consists of two stages: \textit{random-unlearning NWC pruning} and \textit{pruned-to-backdoored OT-based fusion}. In the first stage, we calculate the NWCs based on random unlearning of the backdoored model, and then prune the top-ranking $\gamma$ neurons to eliminate the backdoor effect. In the second stage, we align the weights of the pruned model with those of the backdoored model layer-by-layer using OT, and then fuse them into a single model. This process effectively dilutes the backdoor effect while preserving the clean performance. An illustration of the fusion process is shown in Figure~\ref{fig:overview}.

Our main contributions can be summarized as follows:
\begin{itemize}

  \item We provide a new data-free pruning insight by revealing the positive correlation between NWCs when unlearning random noise and poisoned data. 
    \item We propose a novel data-free defense strategy that combines the high ACC of the backdoored model with the low ASR of the pruned model, using the OT-based model fusion. To our knowledge, this is the first work to apply OT and model fusion techniques to backdoor defense.  
    \item Experiments across various attacks, datasets, and experimental setups validate the effectiveness of our proposed OTBR method. Specifically, OTBR significantly outperforms both state-of-the-art (SOTA) data-free methods and SOTA data-dependent ones, consistently achieving successful defense performance against all tested attacks.
\end{itemize}

\begin{figure*}[t]
\centering
\includegraphics[width=0.9\textwidth]{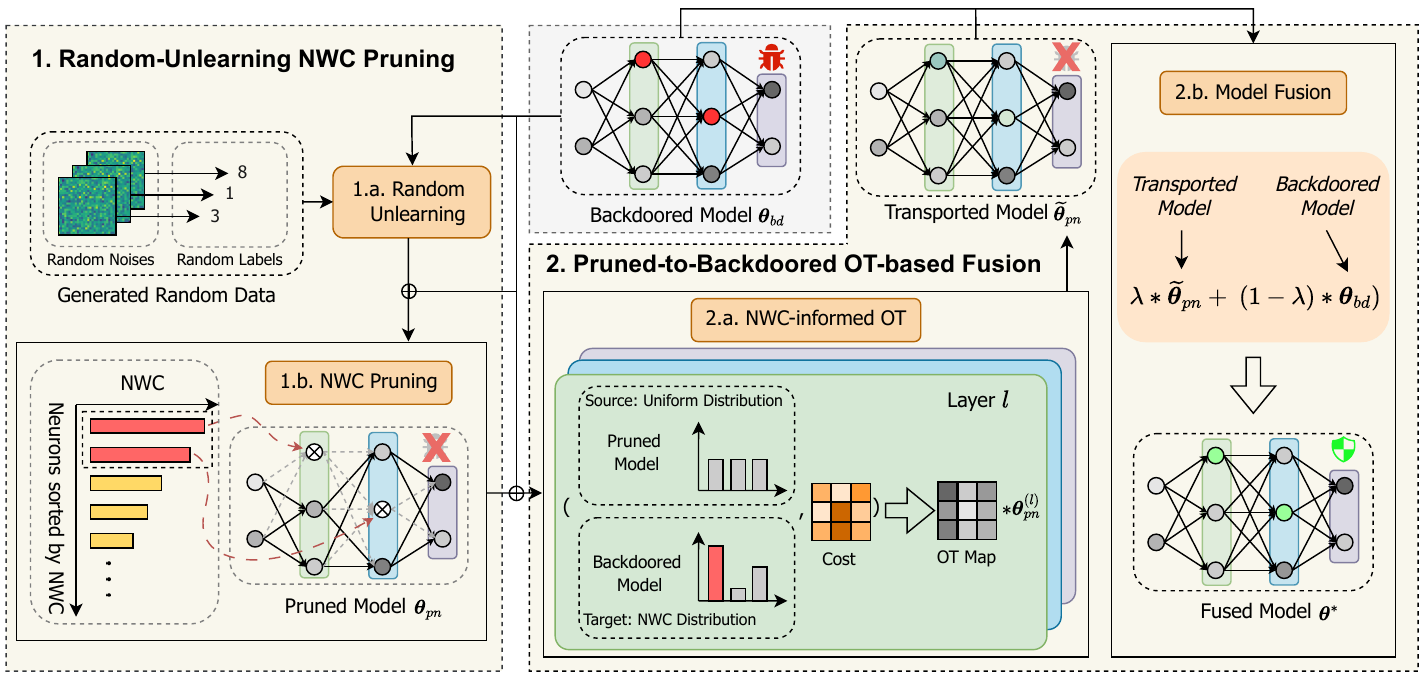} 
\caption{Overview of the proposed \nameFramework framework.}
\label{fig:framework}
\end{figure*}
\section{Related Work}
\label{sec: related work}

\subsection{Backdoor Attack}
In the literature, various backdoor attacks on DNNs have been proposed, which can be generally categorized into two types: data-poisoning attacks and training-controllable attacks. For \textbf{data-poisoning attacks}, adversaries have access to the training dataset.
BadNets~\cite{gu2019badnets}, as one of the earliest examples, was proposed to implant a trigger pattern into the bottom-right corner of a small subset of the training images and reassign the labels to a specific target one. 
To enhance the stealthiness of the trigger, Blended~\cite{chen2017targeted} was proposed to blend the trigger onto the selected data with adjustable opacity. Recently, more sophisticated strategies have been proposed to enhance the trigger, including but not limited to SIG~\cite{barni2019new}, label-consistent attacks~\cite{shafahi2018poison, zhao2020clean}, and SSBA~\cite{li2021invisible}. 
Meanwhile, the second type, \textbf{training-controllable attacks}, is also rapidly evolving. 
In these attacks, adversaries have access to the training process, enabling more advanced attack strategies.  
Representative examples of this category include WaNet~\cite{nguyen2021wanet} and Input-aware~\cite{nguyen2020input}, which incorporate an injection function into the training process to generate unique triggers for each input data. These innovative tactics make it more challenging to detect the triggers and conduct an effective defense.

\subsection{Backdoor Defense}
In general, backdoor defense methods can be categorized into three types\footnote{As the first two categories are less related to the scenario in this paper, the details about them are postponed to the Appendix.}: pre-training, in-training, and post-training defenses.
Among them, \textbf{Post-training Defense} has received the most attention, where the defenders aim to mitigate the backdoor effect of a well-trained backdoored model. FP~\cite{liu2018fine}, as one of the seminal defense methods, prunes the less-activated neurons and then fine-tunes the model, based on the observation that poisoned and clean data activate different neurons; ANP~\cite{wu2021adversarial} detects and prunes backdoor-related neurons by applying adversarial perturbations to neuron weights; Building on this, RNP~\cite{li2023reconstructive} refines the perturbation technique using clean unlearning, and performs pruning based on a learned mask. Except for these pruning-based techniques, there also exist some other important defense techniques. For instance, NC~\cite{wang2019neural} proposes recovering the trigger to improve backdoor removal; NAD~\cite{li2021neural} pioneers the use of model distillation to train a benign student model; i-BAU~\cite{zeng2021adversarial} uses adversarial attacks to identify potential triggers and then performs poison unlearning to mitigate the backdoor effect. 

Different from the above defenses, which are all data-dependent, CLP~\cite{zheng2022data} is the first data-free defense method, which identifies and prunes potential backdoored neurons based on channel Lipschitzness; ABD~\cite{hong2023revisiting} designs a plug-in defensive technique specialized for data-free knowledge distillation; DHBE~\cite{yan2023dhbe} proposes a competing strategy between distillation and backdoor regularization to distill a clean student network without data. 

Although several data-dependent techniques have already been proposed in the literature, the scarcity of data-free defense techniques still limits the applicability of backdoor defenses in real-world scenarios. In this paper, we will focus on addressing this issue, and develop a novel effective data-free defense method by using random-unlearning NWCs and the OT-based model fusion technique. 


\section{Preliminary}
\label{sec:problem}

\subsection{Threat Model}
In this work, we address threats from both data-poisoning and training-controllable attacks.
The attackers aim to poison a small portion of the training data so that the trained model predicts a target class when presented with data containing a pre-defined \textit{trigger}, while otherwise performing normally.
The weights of a $L$-layer backdoored model are denoted as $\boldsymbol{\theta}_{bd}=\{\boldsymbol{\theta}^{(l)}_{bd}\}_{1\leq l\leq L}$, where $\boldsymbol{\theta}^{(l)}_{bd}$ represents the weights for the $l^{th}$ layer, consisting of $m^{(l)}$ neurons.

\subsection{Defense Setting}
We focus on the post-training scenario, aiming to mitigate the backdoor effect of a well-trained backdoored model while minimizing the negative impact on ACC. Different from most previous works~\cite{liu2018fine, wu2021adversarial, zeng2021adversarial}, 
which assumes access to 5\% of clean data for defense, 
we adopt a more stringent approach that relies only on the backdoored model, without access to any clean data~\cite{zheng2022data}.  
\section{Method}
\label{sec:method}

\subsection{Overview of Our Method}
The complete data-free defense process of our proposed \nameFramework strategy is illustrated in Figure~\ref{fig:framework}, which consists of two stages as follows:  

\begin{itemize}
\item In \textbf{Stage 1}, referred to as \textit{random-unlearning NWC pruning}, we aim to obtain a backdoor-free model. Specifically, we first conduct random unlearning on the backdoored model for $I$ iterative steps. During each step, a mini-batch of random noise with random labels is generated and used for unlearning. Then, we calculate the NWC for each neuron based on their weights from both the original backdoored and unlearned models. Finally, we prune the top-ranking $\gamma$ of neurons, based on their NWCs, from the backdoored model to eliminate its backdoor effect. 

\item In \textbf{Stage 2}, referred to as \textit{pruned-to-backdoored OT-based fusion}, when obtaining a sub-optimal pruned model, we aim to combine its low ASR with the high ACC of the original backdoored model by repairing the backdoor-related neurons using OT-based model fusion. Specifically, we propose \textit{NWC-informed OT} to align the weights of the pruned model with those of the backdoored model layer-by-layer, taking into account the backdoor importance as determined by NWCs. For each layer, starting with the earliest pruned one, we initialize the probability mass on neuron weights using a uniform distribution for the pruned model and an NWC distribution for the original backdoored model. This strategy discriminatively transfers clean functionality to the backdoor-related neurons. The cost matrix $\boldsymbol{C}^{(l)}$ is calculated based on the Euclidean distance between neuron weights from the two models. Using this cost matrix, we then derive the optimal transport map $\boldsymbol{\rm T}^{(l)}$ and employ it to transport the weights of the pruned model. After aligning all layers, we perform a simple weight averaging to fuse the transported and backdoored models, resulting in an effective defense. 
\end{itemize}

Next, we will present more detailed formulations and provide further insights. 

\subsection{Stage 1: Random-Unlearning NWC Pruning}

\paragraph{Random Unlearning.}
Unlearning is a reverse training process designed to maximize the loss value on a given dataset~\cite{li2023reconstructive}. 
In this work, we define \textit{random unlearning} as the process of unlearning a DNN model $f$ using a generated random dataset $\mathcal{D}_{r}$.
More precisely, random unlearning on the backdoored model  $\boldsymbol{\theta}_{bd}$ is formulated as:
\begin{equation}
\label{equ:unlearn}
    \max_{\boldsymbol{\theta}_{bd}} \mathbb{E}_{(\boldsymbol{x}_r, y_r) \in \mathcal{D}_r} \left[\mathcal{L}(f(\boldsymbol{x}_r;\boldsymbol{\theta}_{bd}),y_r)\right],
\end{equation}
where the loss function $\mathcal{L}$ is chosen to be a cross-entropy loss, and the generated random dataset $\mathcal{D}_{r}$ contains $I \times B$ pairs of random noises $\boldsymbol{x}_r \in [0,1]^{A\times H\times W}$ and random labels $y_r \in \{0,1,\dots,G\}$.
Here, $I$ is the number of iterative steps; $B$ is the batch size; $A$, $H$ and $W$ represent the generated noise size; and $G$ is the largest class label.  

\paragraph{NWC Pruning.}
We follow the NWC definition from~\cite{lin2024unveiling} to quantify the weight changes for each neuron during unlearning. 
Specifically, for the $j$-th neuron in the $l$-th layer, the NWC is defined as: 
\begin{equation}
\label{equ:nwc}
\mathrm{NWC}^{(l)j} \eqdef \|\boldsymbol{\theta}^{(l)j}_{ul}-\boldsymbol{\theta}^{(l)j}_{bd}\|_1,
\end{equation} 
where $\boldsymbol{\theta}_{ul}$ denotes the unlearned backdoored model, $j \in \{1, \dots, m^{(l)}\}$ and $l \in \{1, \dots, L\}$. 
To eliminate the backdoor effect, we sort all calculated NWCs in descending order and prune the top-ranking $\gamma$ of neurons from the original backdoored model. The pruned model is denoted as $\boldsymbol{\theta}_{pn}$. 

To better understand why substituting clean unlearning with random unlearning is effective for NWC pruning, we provide a possible explanation in the \textbf{Further Analysis of Appendix}. 



\subsection{Stage 2: Pruned-to-Backdoored OT-based Fusion}
\paragraph{Optimal Transport.}
OT is a mathematical framework to find the most economical way to transport mass from one distribution to another. 
Suppose we have two discrete probability distributions in the space $\mathcal{X}=\{x_i\}_{i=1}^n$ and $\mathcal{Y}=\{y_j\}_{j=1}^m$, \ie, the source distribution $\mu := \sum_{i=1}^n \alpha_i \cdot \delta(x_i)$ and the target distribution $\nu := \sum_{j=1}^m \beta_j \cdot \delta(y_j)$, where $\sum_{i=1}^n \alpha_i=\sum_{j=1}^m \beta_j = 1$ and $\delta(\cdot)$ is the Dirac delta function.
The OT problem can be formulated as a linear programming problem as follows: 
\begin{equation}
    \label{equ:ot}
    \begin{aligned}
    & \mathrm{OT}(\mu, \nu ; \boldsymbol{C})\eqdef \min \langle \boldsymbol{\rm T}, \boldsymbol{C}\rangle, \\ 
    & \text { s.t., }  \boldsymbol{\rm T} \mathbf{1}_m=\vect{\alpha}, \ \boldsymbol{\rm T}^{\top} \mathbf{1}_n=\vect{\beta}
    \end{aligned}
    ,
\end{equation}
where $\boldsymbol{\rm T} \in \mathbb{R}_{+}^{n \times m}$ is the transport map that determines the optimal transport amount of mass from $\mathcal{X}$ to $\mathcal{Y}$, and $\boldsymbol{C}$ is the cost matrix quantifying the cost of moving each unit of mass. 

\paragraph{NWC-informed OT.}
In our approach, we align the NWC-pruned model, which contains only clean functionality, with the original backdoored model to achieve more effective fusion. The goal is to dilute the backdoor effect with the least influence on clean performance. To achieve this, we focus more on the backdoor-related neurons during weight transport by employing NWC-informed initialization for the target distribution $\nu$ (backdoored model), while using a uniform distribution for the source distribution $\mu$ (pruned model). 
For the $l$-th layer, we denote the probability mass as:  
\begin{equation}
    \label{equ:prob}
    \vect{\alpha}^{(l)} \eqdef \left\{\frac{1}{n^{(l)}}\right\}_{i=1}^{n^{(l)}} , \ 
    \vect{\beta}^{(l)} \eqdef \left\{ \frac{\mathrm{NWC}^{(l)j}}{\sum_{j=1}^{m^{(l)}}\mathrm{NWC}^{(l)j}}\right\}_{j=1}^{m^{(l)}},
\end{equation}
where $n^{(l)}$ and $m^{(l)}$ denote the neuron numbers of pruned and backdoored models, respectively.  
Then, based on the distributions $\mu^{(l)}$ and $\nu^{(l)}$, and the cost matrix $\boldsymbol{C}^{(l)}$, we can derive the optimal transport map $\boldsymbol{\rm T}^{(l)}$ by equation~\eqref{equ:ot}.

\paragraph{Model Fusion.}
 
Inspired by the OTFusion~\cite{singh2020model}, we align and fuse the pruned and backdoored models layer-by-layer, using OT in equation \eqref{equ:ot} and our defined distributions in equation \eqref{equ:prob}. The details of the entire fusion process are shown in Algorithm~\ref{alg:otfusion}.  
Note that we start the fusion process from the first pruned layer $p$, rather than the second layer. 
For the $l$-th layer, $n^{(l)}$ and $m^{(l)}$ represent the neuron number of $\boldsymbol{\theta}_{pn}^{(l)}$ and $\boldsymbol{\theta}_{bd}^{(l)}$, respectively.

In Algorithm~\ref{alg:otfusion}, for each layer $l$, we first align the incoming edge weights using the OT map $\boldsymbol{\rm T}^{(l-1)}$ and probability mass $\vect{\beta}^{(l-1)}$ from the previous layer:
\begin{equation*}
    \widehat{\boldsymbol{\theta}}_{pn}^{(l)} \leftarrow 
        \boldsymbol{\theta}_{pn}^{(l)} \boldsymbol{\rm T}^{(l-1)}\textrm{diag}(1/\vect{\beta}^{(l-1)}).
\end{equation*} 
Then, we get the distributions $\mu^{(l)}$ and $\nu^{(l)}$ of the current layer and compute the cost matrix $\boldsymbol{C}^{(l)}$ using Euclidean distance between neuron weights: $\boldsymbol{C}^{(l)}_{ij} \eqdef  \|\boldsymbol{\theta}_{pn}^{(l)i}-\boldsymbol{\theta}_{bd}^{(l)j} \|^2.$ 
Finally, as in equation \eqref{equ:ot}, using $\mu^{(l)}$, $\nu^{(l)}$ and $\boldsymbol{C}^{(l)}$, the OT map $\boldsymbol{\rm T}^{(l)}$ of the current layer can be derived, and the transported pruned model $\widetilde{\boldsymbol{\theta}}_{pn}^{(l)}$ can be obtained as:
\begin{equation*}
    \label{equ:align}
    \widetilde{\boldsymbol{\theta}}_{pn}^{(l)} \leftarrow \textrm{diag}\left(1/\vect{\beta}^{(l)}\right)\boldsymbol{\rm T}^{(l)^\top}\boldsymbol{\widehat{\theta}}_{pn}^{(l)}, 
\end{equation*}
which has been aligned with the backdoored model. 

After alignment, the transported model is then fused with the backdoored model to obtain the final defense model, which can be formulated as:
\begin{equation*}
    \label{equ:fusion}
    \boldsymbol{\theta}^* \leftarrow \lambda  \widetilde{\boldsymbol{\theta}}_{pn} + (1- \lambda)\boldsymbol{\theta}_{bd},
\end{equation*}
where $\boldsymbol{\theta}^*$ represents the fused clean model and $\lambda$ is the balance coefficient.




\begin{algorithm}[t]
    \caption{Pruned-to-Backdoored OT-based Fusion}
    \label{alg:otfusion}
    \raggedright
    {\bf Input}: Pruned model $\boldsymbol{\theta}_{pn}$, backdoored model $\boldsymbol{\theta}_{bd}$, random-unlearning NWC values for each neuron, balance coefficient $\lambda$, the first pruned layer $p$.\\
    {\bf Output}: Clean model $\boldsymbol{\theta}^*$.\\
    \begin{algorithmic} [1]
    \STATE $\vect{\alpha}^{(p-1)} \leftarrow \left\{1/n^{(p-1)}\right\}_{i=1}^{n^{(p-1)}}$ \\
    $\vect{\beta}^{(p-1)} \leftarrow \left\{1/m^{(p-1)}\right\}_{j=1}^{m^{(p-1)}}$
    \STATE $\boldsymbol{\rm T}^{(p-1)} \leftarrow \textrm{diag}(\vect{\beta}^{(p-1)})\mathbf{I}_{m^{(p-1)}\times m^{(p-1)}}$
    \FOR{$l=p\ \TO\ L$}
        \STATE $\widehat{\boldsymbol{\theta}}_{pn}^{(l)} \leftarrow 
        \boldsymbol{\theta}_{pn}^{(l)} \boldsymbol{\rm T}^{(l-1)}\textrm{diag}(1/\vect{\beta}^{(l-1)})$
        \STATE $\vect{\alpha}^{(l)} \leftarrow \left\{1/n^{(l)}\right\}_{i=1}^{n^{(l)}} $
        \STATE $\vect{\beta}^{(l)} \leftarrow \left\{ \mathrm{NWC}^{(l)j}/\sum_{j=1}^{m^{(l)}}\mathrm{NWC}^{(l)j}\right\}_{j=1}^{m^{(l)}}$
        \STATE $\mu^{(l)}, \nu^{(l)} \leftarrow \textrm{GetDistribution}(\vect{\alpha}^{(l)}, \vect{\beta}^{(l)})$
        \STATE $\boldsymbol{C}^{(l)} \leftarrow \textrm{ComputeCost}(\boldsymbol{\theta}_{pn}^{(l)}, \boldsymbol{\theta}_{bd}^{(l)})$
        \STATE $\boldsymbol{\rm T}^{(l)} \leftarrow \textrm{OT}(\mu^{(l)},\nu^{(l)}, \boldsymbol{C}^{(l)})$
        \STATE $\widetilde{\boldsymbol{\theta}}_{pn}^{(l)} \leftarrow \textrm{diag}(1/\vect{\beta}^{(l)})\boldsymbol{\rm T}^{(l)^\top}\widehat{\boldsymbol{\theta}}_{pn}^{(l)}$
        \STATE $\boldsymbol{\theta}^{*(l)} \leftarrow \lambda  \widetilde{\boldsymbol{\theta}}_{pn}^{(l)} + (1- \lambda)\boldsymbol{\theta}_{bd}^{(l)}$
    \ENDFOR
    \STATE Obtain clean model $\boldsymbol{\theta}^{*}$
    \end{algorithmic}
\end{algorithm}

\paragraph{Why Can OT-based Fusion Mitigate Backdoor Effect?}
We now offer a possible explanation for the effectiveness of OT-based model fusion in mitigating backdoor effects.  
Based on previous work~\cite{lin2024unveiling}, the NWC-pruned model can be made backdoor-free, \ie, the ASR dropping to zero, by selecting a suitable pruning threshold. Therefore, by aligning the pruned model with the original backdoored model using NWC-informed OT, we can transport the clean functionality of the remaining neurons to the nearest backdoored positions, as determined by NWCs and Euclidean distance. Then, further fusion based on the transported model can be viewed as a dilution operation to weaken the backdoored effect of the original backdoored model while preserving its clean functionality, thanks to the inherent ability of OT~\cite{singh2020model, theus2024towards}. This is consistent with the previous insights that the backdoor task is easier and encoded in much fewer neurons than the clean task~\cite{li2021anti, cai2022randomized}.

\begin{figure}[]
\centering
\includegraphics[width=0.7\linewidth]{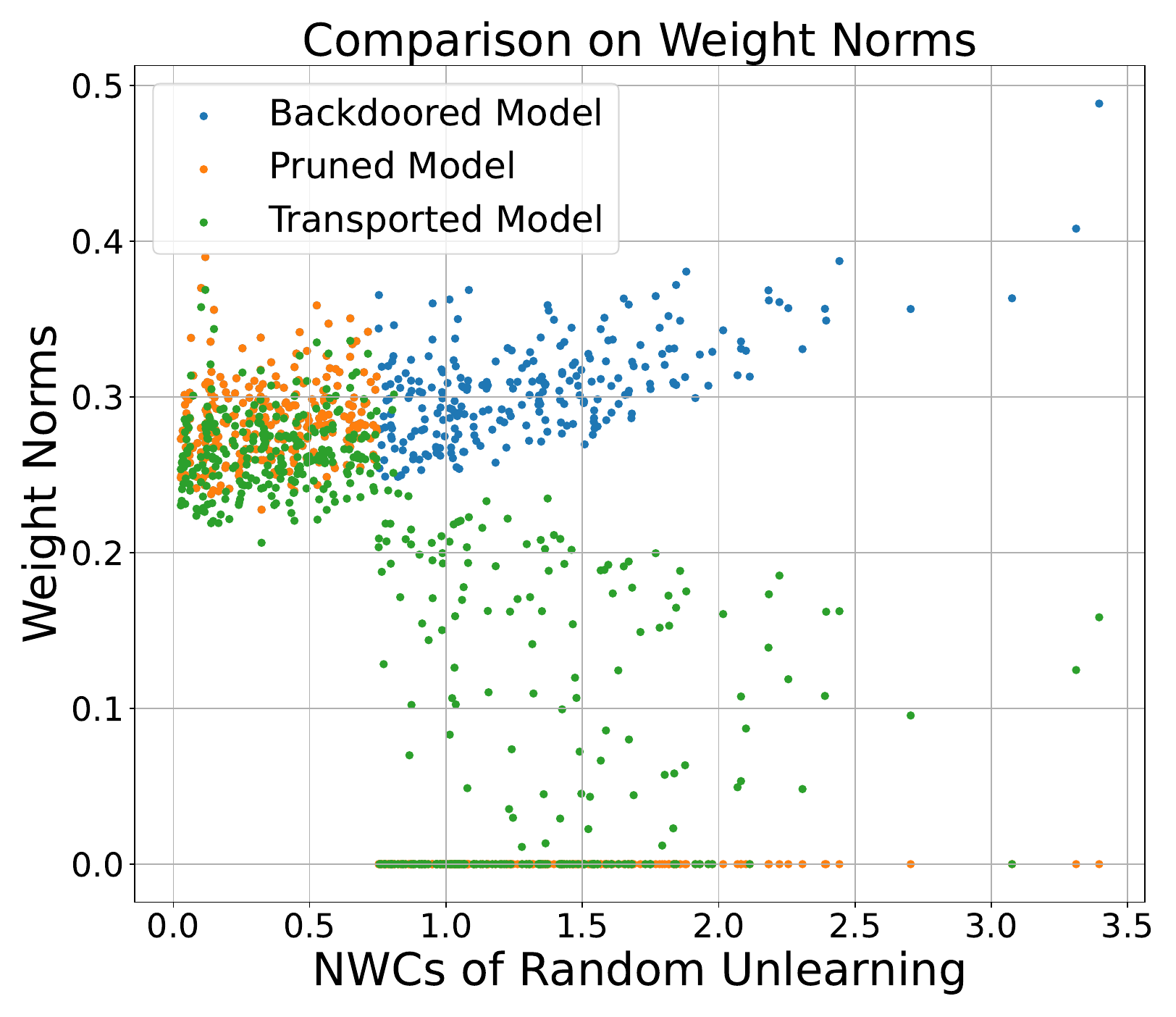} 
\caption{Illustration of neuron-level weight norms for the backdoored, pruned, and transported models during OT-based fusion. }
\label{fig:fusion}
\end{figure}

A practical example of the BadNets-attacked PreAct-ResNet18~\cite{he2016identity} is illustrated in Figure~\ref{fig:fusion}. 
From the perspective of \textit{weight norm},
the larger the difference in a neuron's weight between the pruned and transported models, the more it is transported by the OT. We observe a consistent, slight decrease in the unpruned neurons that are intensively transported to some specific pruned neurons, resulting in their rapid recovery. This outcome reflects the effect of transporting from a uniform source distribution to a NWC-informed target distribution. By fusing the transported (green) and backdoored (blue) models, we can discriminately modify the neuron functionality, \eg, recovering more in low-NWC neurons, to effectively mitigate the backdoor effect while preserving high performance.
More observations on activations are provided in the \textbf{Further Analysis of Appendix}. 


\section{Experiment}
\label{sec:experiment}

\begin{table*}[]
\caption{Performance comparison with the SOTA defenses on CIFAR-10, Tiny ImageNet, and CIFAR-100 (\%).}
\label{tab:main_perform}
\centering
\resizebox{1\linewidth}{!}{
\begin{tabular}{c|c|cc|cccccccccc|cccc}
\hline
\multirow{3}{*}{Datasets}      & \multirow{3}{*}{Attacks} & \multicolumn{2}{c|}{\multirow{2}{*}{No Defense}} & \multicolumn{10}{c|}{\textbf{Data-Dependent}}                                                                                                                                         & \multicolumn{4}{c}{\textbf{Data-Free}}                                  \\ \cline{5-18} 
                               &                          & \multicolumn{2}{c|}{}                            & \multicolumn{2}{c|}{FP}            & \multicolumn{2}{c|}{NAD}           & \multicolumn{2}{c|}{ANP}           & \multicolumn{2}{c|}{i-BAU}         & \multicolumn{2}{c|}{RNP} & \multicolumn{2}{c|}{CLP}            & \multicolumn{2}{c}{\textbf{\nameFramework}} \\ \cline{3-18} 
                               &                          & ACC                    & ASR                     & ACC   & \multicolumn{1}{c|}{ASR}   & ACC   & \multicolumn{1}{c|}{ASR}   & ACC   & \multicolumn{1}{c|}{ASR}   & ACC   & \multicolumn{1}{c|}{ASR}   & ACC         & ASR        & ACC   & \multicolumn{1}{c|}{ASR}    & ACC         & ASR        \\ \hline
\multirow{8}{*}{CIFAR-10}      & BadNets                  & 91.32                  & 95.03                   & 91.31 & \multicolumn{1}{c|}{57.13} & \greencell{89.87} & \multicolumn{1}{c|}{\greencell{2.14}}  & \greencell{90.94} & \multicolumn{1}{c|}{\greencell{5.91}}  & \greencell{89.15} & \multicolumn{1}{c|}{\greencell{1.21}}  & 89.81       & 24.97      & 90.06 & \multicolumn{1}{c|}{77.50}  & \greencell{90.11}       & \greencell{1.08}       \\
                               & Blended                  & 93.47                  & 99.92                   & 93.17 & \multicolumn{1}{c|}{99.26} & 92.17 & \multicolumn{1}{c|}{97.69} & 93.00 & \multicolumn{1}{c|}{84.90} & 87.00 & \multicolumn{1}{c|}{50.53} & 88.76       & 79.74      & 91.32 & \multicolumn{1}{c|}{99.74}  & \greencell{92.01}       & \greencell{1.64}       \\
                               & Input-aware              & 90.67                  & 98.26                   & \greencell{91.74} & \multicolumn{1}{c|}{\greencell{0.04}}  & \greencell{93.18} & \multicolumn{1}{c|}{\greencell{1.68}}  & \greencell{91.04} & \multicolumn{1}{c|}{\greencell{1.32}}  & 89.17 & \multicolumn{1}{c|}{27.08} & \greencell{90.52}       & \greencell{1.84}       & \greencell{90.30} & \multicolumn{1}{c|}{\greencell{2.17}}   & \greencell{86.52}       & \greencell{0.37}       \\
                               & LF                       & 93.19                  & 99.28                   & 92.90 & \multicolumn{1}{c|}{98.97} & 92.37 & \multicolumn{1}{c|}{47.83} & 92.83 & \multicolumn{1}{c|}{54.99} & 84.36 & \multicolumn{1}{c|}{44.96} & \greencell{88.43}       & \greencell{7.02}       & 92.84 & \multicolumn{1}{c|}{99.18}  & \greencell{87.68}       & \greencell{9.69}       \\
                               & SSBA                     & 92.88                  & 97.86                   & 92.54 & \multicolumn{1}{c|}{83.50} & 91.91 & \multicolumn{1}{c|}{77.40} & 92.67 & \multicolumn{1}{c|}{60.16} & \greencell{87.67} & \multicolumn{1}{c|}{\greencell{3.97}}  & \greencell{88.60}       & \greencell{17.89}      & 91.38 & \multicolumn{1}{c|}{68.13}  & \greencell{85.27}       & \greencell{9.54}       \\
                               & Trojan                   & 93.42                  & 100.00                  & 92.46 & \multicolumn{1}{c|}{71.17} & \greencell{91.88} & \multicolumn{1}{c|}{\greencell{3.73}}  & 92.97 & \multicolumn{1}{c|}{46.27} & \greencell{90.37} & \multicolumn{1}{c|}{\greencell{2.91}}  & \greencell{90.89}       & \greencell{3.59}       & 92.98 & \multicolumn{1}{c|}{100.00} & \greencell{90.62}       & \greencell{7.50}       \\
                               & WaNet                    & 91.25                  & 89.73                   & \greencell{91.46} & \multicolumn{1}{c|}{\greencell{1.09}}  & 93.17 & \multicolumn{1}{c|}{22.98} & \greencell{91.32} & \multicolumn{1}{c|}{\greencell{2.22}}  & \greencell{89.49} & \multicolumn{1}{c|}{\greencell{5.21}}  & \greencell{90.43}       & \greencell{0.96}       & 81.91 & \multicolumn{1}{c|}{78.42}  & \greencell{88.12}       & \greencell{10.93}      \\ \cline{2-18} 
                               & Average                  & 92.31                  & 97.15                   & \textbf{92.23} & \multicolumn{1}{c|}{58.74} & 92.08 & \multicolumn{1}{c|}{36.21} & 92.11 & \multicolumn{1}{c|}{36.54} & 88.17 & \multicolumn{1}{c|}{19.41} & 89.63       & 19.43      & 90.11 & \multicolumn{1}{c|}{75.02}  & 88.62       & \textbf{5.82}       \\ \hline
\multirow{6}{*}{\begin{tabular}[c]{@{}c@{}}Tiny \\ ImageNet\end{tabular}} & BadNets                  & 56.23                  & 100.00                  & 51.73 & \multicolumn{1}{c|}{99.99} & \greencell{46.37} & \multicolumn{1}{c|}{\greencell{0.27}}  & \greencell{50.55} & \multicolumn{1}{c|}{\greencell{7.74}}  & 51.48 & \multicolumn{1}{c|}{97.36} & 21.91       & 0.00       & 55.94 & \multicolumn{1}{c|}{100.00} & \greencell{54.13}       & \greencell{0.00}       \\
                               & Input-aware              & 57.45                  & 98.85                   & 55.28 & \multicolumn{1}{c|}{62.92} & \greencell{47.91} & \multicolumn{1}{c|}{\greencell{1.86}}  & \greencell{53.17} & \multicolumn{1}{c|}{\greencell{0.17}}  & 52.48 & \multicolumn{1}{c|}{72.98} & 15.57       & 0.00       & 57.75 & \multicolumn{1}{c|}{99.58}  & \greencell{51.40}       & \greencell{0.02}       \\
                               & SSBA                     & 55.22                  & 97.71                   & 50.47 & \multicolumn{1}{c|}{88.87} & 45.32 & \multicolumn{1}{c|}{57.32} & 52.83 & \multicolumn{1}{c|}{91.44} & 49.86 & \multicolumn{1}{c|}{81.90} & 37.64       & 0.00       & 55.17 & \multicolumn{1}{c|}{97.65}  & \greencell{54.15}       & \greencell{3.89}       \\
                               & Trojan                   & 55.89                  & 99.98                   & \greencell{50.22} & \multicolumn{1}{c|}{\greencell{8.82}}  & \greencell{48.48} & \multicolumn{1}{c|}{\greencell{0.83}}  & \greencell{50.37} & \multicolumn{1}{c|}{\greencell{1.40}}  & 52.65 & \multicolumn{1}{c|}{98.49} & \greencell{46.27}       & \greencell{0.00}       & \greencell{55.86} & \multicolumn{1}{c|}{\greencell{8.39}}   & \greencell{53.85}       & \greencell{0.14}       \\
                               & WaNet                    & 56.78                  & 99.49                   & \greencell{53.84} & \multicolumn{1}{c|}{\greencell{3.94}}  & \greencell{46.98} & \multicolumn{1}{c|}{\greencell{0.43}}  & \greencell{53.87} & \multicolumn{1}{c|}{\greencell{0.75}}  & 53.71 & \multicolumn{1}{c|}{75.23} & 20.50       & 0.00       & 56.21 & \multicolumn{1}{c|}{98.50}  & \greencell{55.64}       & \greencell{0.03}       \\ \cline{2-18} 
                               & Average                  & 56.31                  & 99.21                   & 52.31 & \multicolumn{1}{c|}{52.91} & 47.01 & \multicolumn{1}{c|}{12.14} & 52.16 & \multicolumn{1}{c|}{20.30} & 52.04 & \multicolumn{1}{c|}{85.19} & 28.38       & \textbf{0.00}       & \textbf{56.19} & \multicolumn{1}{c|}{80.82}  & 53.83       & 0.82       \\ \hline
\multirow{5}{*}{CIFAR-100}     & BadNets                  & 67.22                  & 87.43                   & \greencell{64.55} & \multicolumn{1}{c|}{\greencell{0.42}}  & \greencell{66.37} & \multicolumn{1}{c|}{\greencell{0.06}}  & \greencell{63.65} & \multicolumn{1}{c|}{\greencell{0.00}}  & \greencell{60.37} & \multicolumn{1}{c|}{\greencell{0.04}}  & 55.68       & 0.00       & 65.40 & \multicolumn{1}{c|}{81.95}  & \greencell{66.81}      & \greencell{0.00}       \\
                               & Input-aware              & 65.24                  & 98.61                   & \greencell{67.82} & \multicolumn{1}{c|}{\greencell{2.34}}  & 69.25 & \multicolumn{1}{c|}{31.11} & \greencell{58.99} & \multicolumn{1}{c|}{\greencell{0.00}}  & 65.21 & \multicolumn{1}{c|}{85.14} & \greencell{55.66}       & \greencell{0.01}       & 65.22 & \multicolumn{1}{c|}{99.81}  & \greencell{59.64}       & \greencell{4.21}       \\
                               & SSBA                     & 69.06                  & 97.22                   & \greencell{61.60} & \multicolumn{1}{c|}{\greencell{14.02}} & 67.38 & \multicolumn{1}{c|}{89.51} & 64.35 & \multicolumn{1}{c|}{39.60} & 63.09 & \multicolumn{1}{c|}{28.91} & 68.44       & 92.80      & 65.39 & \multicolumn{1}{c|}{97.52}  & \greencell{66.89}       & \greencell{1.28}       \\
                               & WaNet                    & 64.04                  & 97.72                   & \greencell{68.07} & \multicolumn{1}{c|}{\greencell{10.29}} & \greencell{68.46} & \multicolumn{1}{c|}{\greencell{0.55}}  & \greencell{60.05} & \multicolumn{1}{c|}{\greencell{0.05}}  & 65.31 & \multicolumn{1}{c|}{43.96} & 49.48       & 0.00       & 25.90 & \multicolumn{1}{c|}{83.49}  & \greencell{62.91}       & \greencell{8.18}       \\ \cline{2-18} 
                               & Average                  & 66.39                  & 95.25                   & 65.51 & \multicolumn{1}{c|}{6.77}  & \textbf{67.87} & \multicolumn{1}{c|}{30.31} & 61.76 & \multicolumn{1}{c|}{9.91}  & 63.50 & \multicolumn{1}{c|}{39.51} & 57.32       & 23.20      & 55.48 & \multicolumn{1}{c|}{90.69}  & 64.06       & \textbf{3.42}       \\ \hline
\multicolumn{4}{c|}{Average ACC Drop (smaller is better)}                                                                                                                                                             & \textbf{$\downarrow$1.51}  & \multicolumn{1}{c|}{-}      & $\downarrow$2.64  & \multicolumn{1}{c|}{-}      & $\downarrow$2.55  & \multicolumn{1}{c|}{-}      & $\downarrow$3.87  & \multicolumn{1}{c|}{-}      & $\downarrow$12.17         &     -        & $\downarrow$3.73  & \multicolumn{1}{c|}{-}       & $\downarrow$2.97          &      -       \\ \hline
\multicolumn{4}{c|}{Average ASR Drop (larger is better)}                                                                                                                                                             &      - & \multicolumn{1}{c|}{$\downarrow$53.39} &    -   & \multicolumn{1}{c|}{$\downarrow$70.11} &   -    & \multicolumn{1}{c|}{$\downarrow$72.51} &  -     & \multicolumn{1}{c|}{$\downarrow$52.33} &       -        & $\downarrow$83.02       &    -   & \multicolumn{1}{c|}{$\downarrow$16.57}  &       -        & $\downarrow$\textbf{93.66}       \\ \hline
\multicolumn{4}{c|}{Successful Defense Count}                                                                                                                                                     & \multicolumn{2}{c|}{8 / 16}        & \multicolumn{2}{c|}{9 / 16}        & \multicolumn{2}{c|}{10 / 16}       & \multicolumn{2}{c|}{5 / 16}        & \multicolumn{2}{c|}{7 / 16} & \multicolumn{2}{c|}{2 / 16}         & \multicolumn{2}{c}{\textbf{16} / 16} \\ \hline
\end{tabular}}
\end{table*}

\subsection{Experimental Setup}
For a fair comparison, all experiments, including the code implementation of our proposed method, are conducted using the default settings in BackdoorBench~\cite{wu2022backdoorbench}.

\paragraph{Datasets.} 
Similar to previous works~\cite{zhu2023enhancing, wei2023shared}, our experiments are conducted on three benchmark datasets, including CIFAR-10~\cite{krizhevsky2009learning}, Tiny ImageNet~\cite{le2015tiny}, and CIFAR-100~\cite{krizhevsky2009learning}.

\paragraph{Attack Setup.}
We evaluate the effectiveness of all defense methods using seven SOTA backdoor attacks: BadNets~\cite{gu2019badnets}, Blended~\cite{chen2017targeted}, Input-aware~\cite{nguyen2020input}, LF~\cite{zeng2021rethinking}, SSBA~\cite{li2021invisible}, Trojan~\cite{liu2018trojaning} and WaNet~\cite{nguyen2021wanet}. All attacks are conducted using the default settings in BackdoorBench~\cite{wu2022backdoorbench}. For example, we set the target label to $0$, the poisoning ratio to $10\%$, and the tested model to PreAct-ResNet18~\cite{he2016identity}.

\paragraph{Defense Setup.}
We compare our proposed \nameFramework method with six SOTA defense methods: Fine-pruning (FP)~\cite{liu2018fine}, NAD~\cite{li2021neural}, ANP~\cite{wu2021adversarial}, i-BAU~\cite{zeng2021adversarial}, RNP~\cite{li2023reconstructive}, and CLP~\cite{zheng2022data}. Note that only CLP can be conducted in a data-free manner similar to \nameFramework, while the other five defenses are all data-dependent. Therefore, we follow the common setting in the post-training scenario that 5\% clean data is provided for those methods. 
Due to space limit, more detailed settings regarding our method are postponed to the \textbf{Implementation Details of Appendix}. 

\paragraph{Evaluation Metrics.}
We use two common metrics to evaluate performance: ACC and ASR.  
They measure the proportion of correct predictions on clean data (the higher, the better) and the rate of incorrect predictions for the target label on poisoned data (the lower, the better), respectively. 
A defense is usually considered successful against an attack if the ASR is reduced to below 20\%~\cite{qi2023towards,xiebadexpert}. 
In this paper, to consider both ACC and ASR, we consider a defense to be \textbf{successful} (marked with \colorbox{successGreen}{green} in all tables) only if it achieves both of the following criteria: ACC decreases by less than 10\% and ASR falls below 20\%. Otherwise, it is considered unsuccessful. The best average results are \textbf{boldfaced} in all tables within this section. 

\subsection{Compared with Previous Works}
The main defense performance, compared with the six baseline methods, is shown in Table~\ref{tab:main_perform}. We observe that our \nameFramework successfully defends against all 16 attacks across three benchmark datasets, achieving the largest drop in average ASR (93.66\%) with an acceptable average ACC reduction (2.97\%). Moreover, \nameFramework outperforms both data-free and data-dependent defenses, achieving the lowest average ASR on CIFAR-10 and CIFAR-100, and the second-best ASR on Tiny ImageNet. Notably, the best ASR on Tiny ImageNet, achieved by RNP, comes with a significant drop in ACC. For the performances of baseline methods, we observe that the data-dependent approaches have clear advantages over the data-free CLP, which succeeds in only 2 out of 16 defenses. ANP achieves the best results with 10 out of 16 successful defenses; however, it consistently fails against SSBA attacks, a shortcoming also observed in NAD. Despite its failures on CIFAR-10, FP performs well on CIFAR-100 and consistently achieves high ACCs across all three datasets, validating the effectiveness of fine-tuning. i-BAU fails completely on Tiny ImageNet, exposing its limitations when dealing with different data complexities. Although RNP achieves good performance in ASR, it fails with significant drops in ACC on Tiny ImageNet. In contrast, \nameFramework performs the best, consistently achieving superior results across various attacks and datasets. 
For a comprehensive evaluation, we also discuss the computational cost, the impact of different poisoning ratios, and the performance on different models in the \textbf{More Experiments of Appendix}.

\subsection{Ablation Studies}

\begin{table}[]
\caption{Comparison of different fusion schemes on CIFAR-10 (\%). $\textbf{V}_1$: no fusion; $\textbf{V}_2$: vanilla fusion; $\textbf{V}_3$: OT-based fusion.}
\label{tab:abl_fusion}
\centering
\resizebox{0.9\linewidth}{!}{
\begin{tabular}{c|cc|cc|cc}
\hline
\multirow{2}{*}{Attacks} & \multicolumn{2}{c|}{$\textbf{V}_1$} & \multicolumn{2}{c|}{$\textbf{V}_2$} & \multicolumn{2}{c}{$\textbf{V}_3$ (Ours)} \\ \cline{2-7} 
                         & ACC                 & ASR                & ACC              & ASR              & ACC            & ASR          \\ \hline
BadNets                  & \greencell{84.98}               & \greencell{1.56}               & 91.3             & 81.13            & \greencell{90.11}          & \greencell{1.08}         \\
Blended                  & 69.08               & 4.71               & \greencell{92.06}            & \greencell{16.76}            & \greencell{92.01}          & \greencell{1.64}         \\
LF                       & 54.15               & 0.83               & 90.97            & 60.4             & \greencell{87.68}          & \greencell{9.69}         \\
SSBA                     & 40.58               & 0.02               & 89.84            & 40.57            & \greencell{85.27}          & \greencell{9.54}         \\ \hline
\end{tabular}}
\end{table}

\begin{figure*}[t]
\centering
\includegraphics[width=0.24\textwidth]{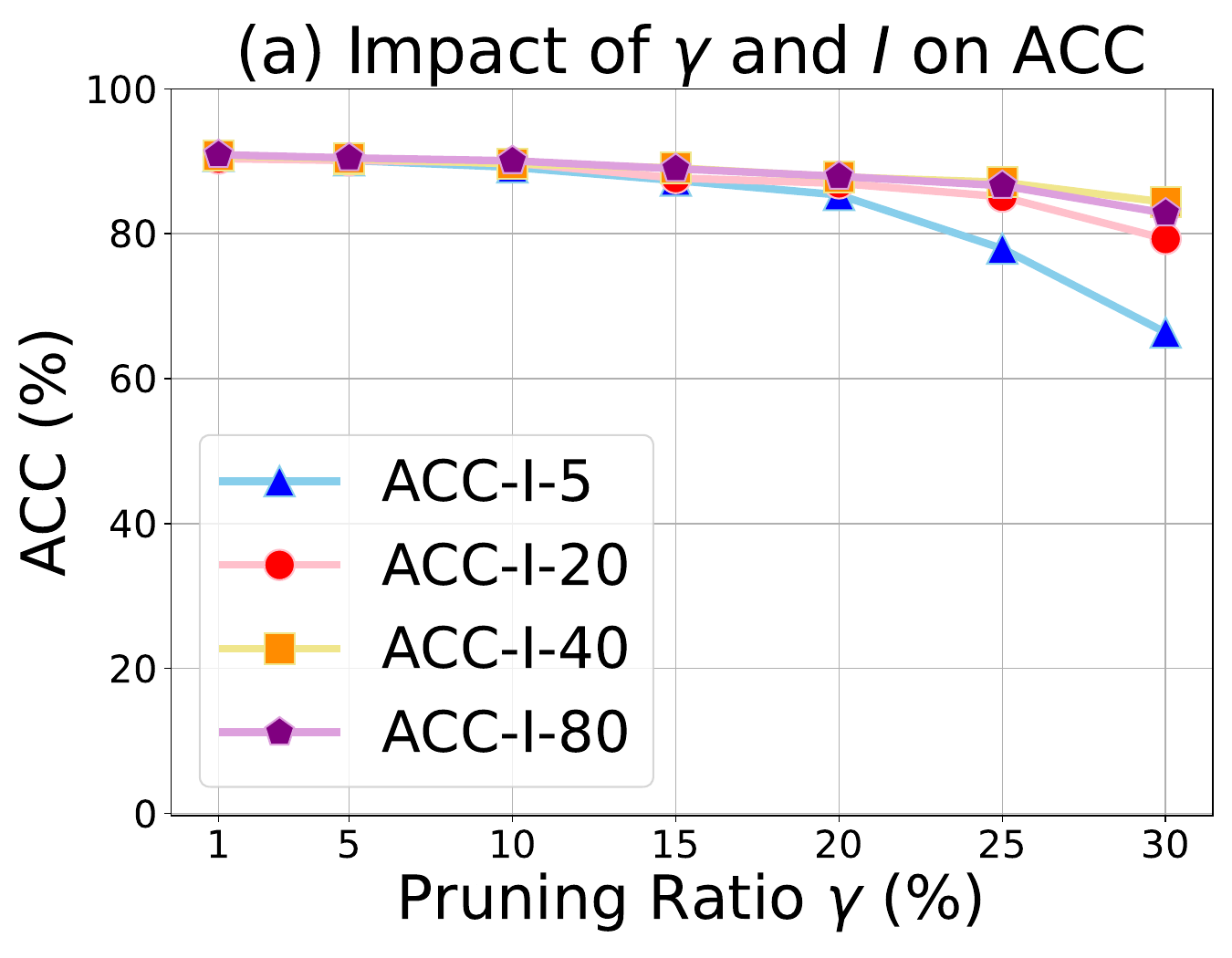} 
\includegraphics[width=0.24\textwidth]{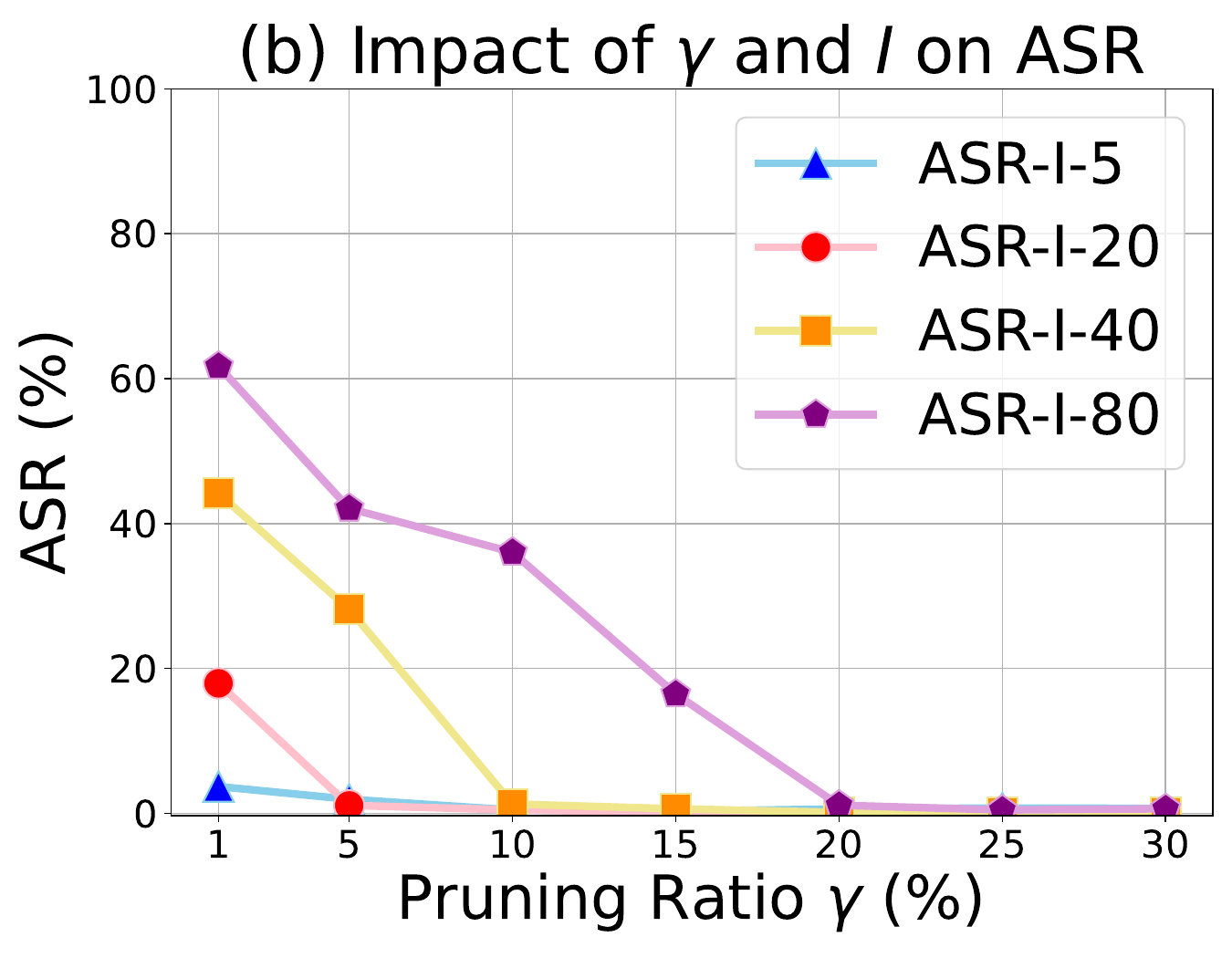}
\includegraphics[width=0.24\textwidth]{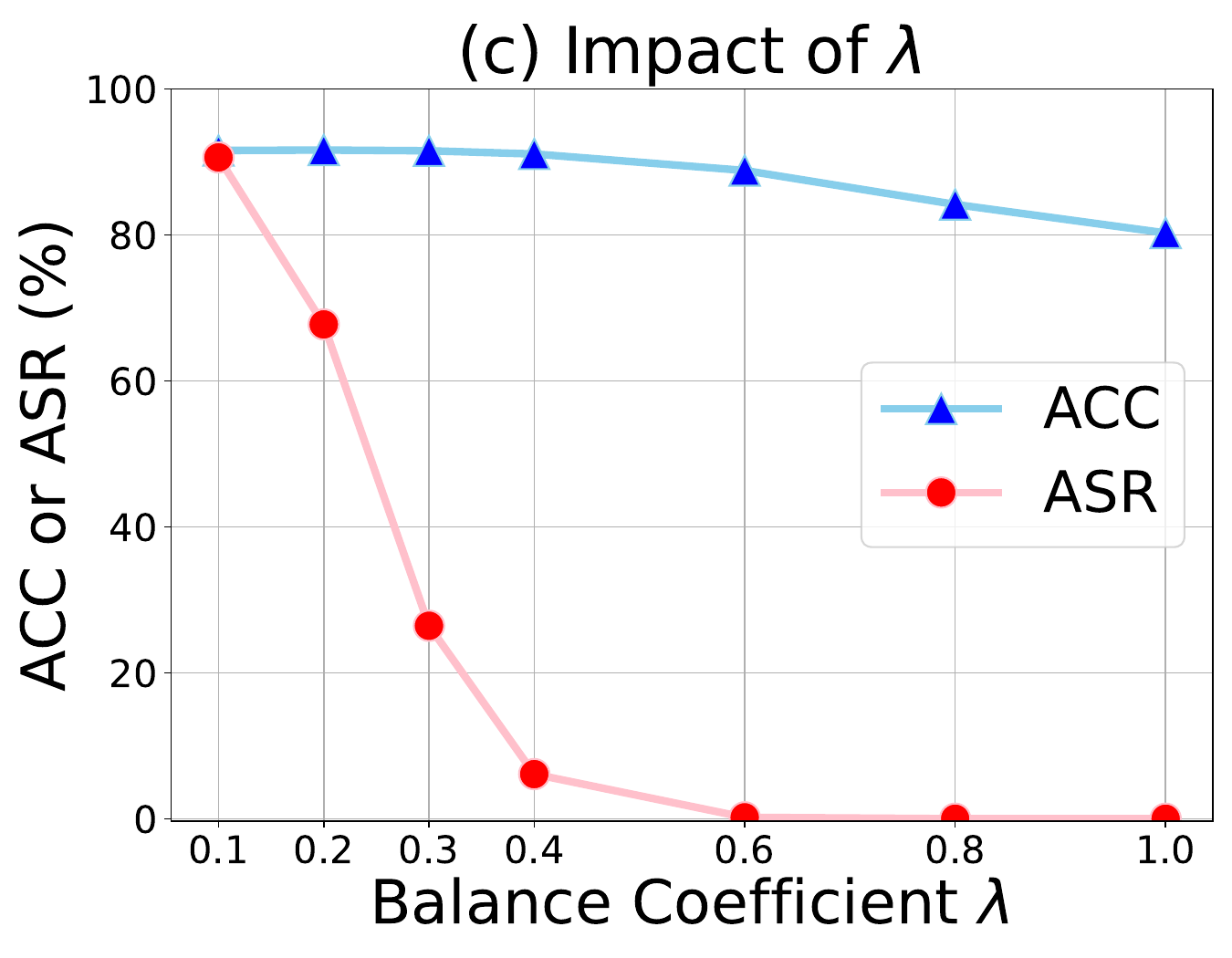}
\includegraphics[width=0.24\textwidth]{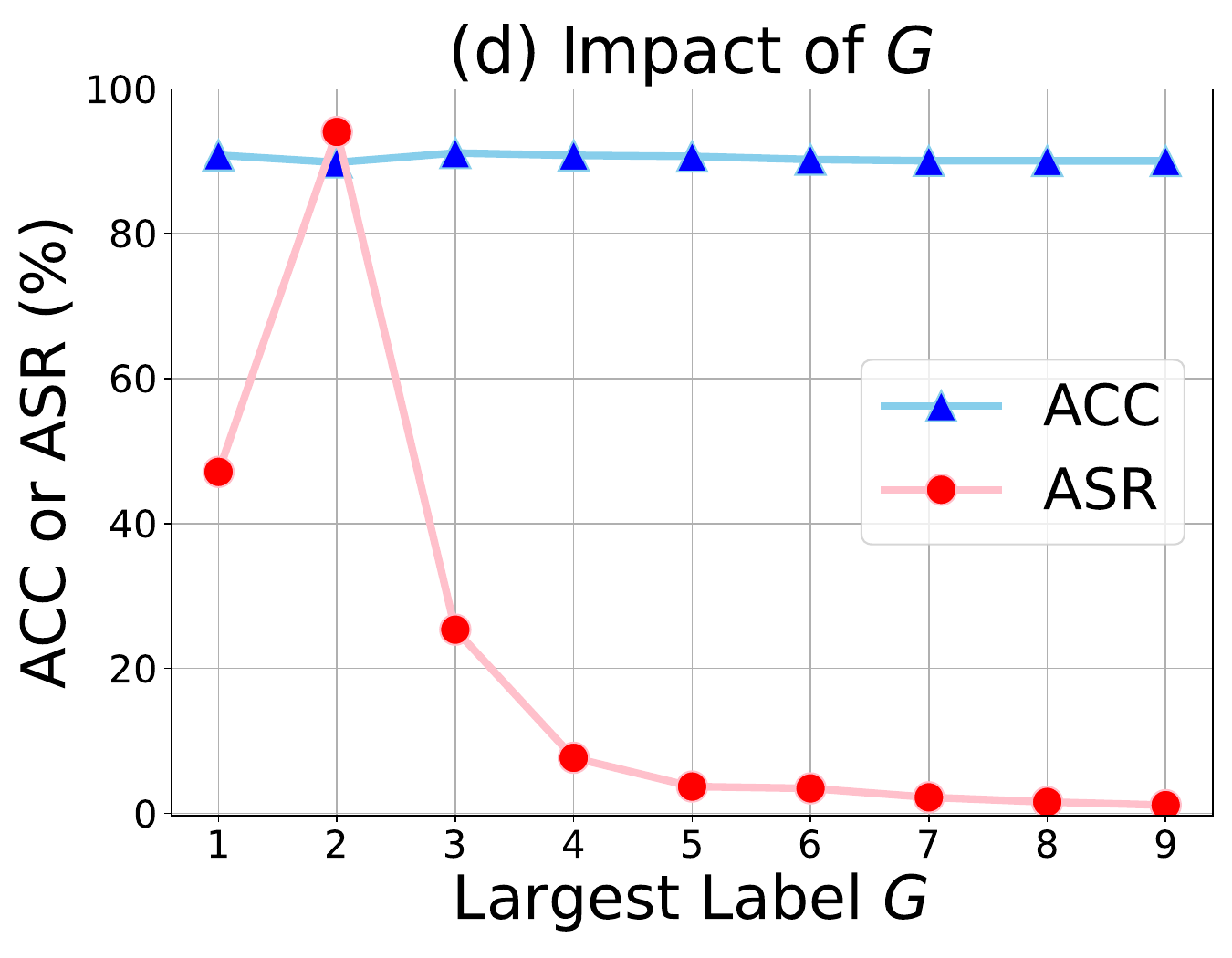}
\caption{Impact of different factors on performance. (a) and (b) show the impact of $\gamma$ and $I$ on ACC and ASR, respectively, with ``ACC-I-5'' representing ACC when $I=5$; (c) shows the impact of $\lambda$; and (d) shows the impact of $G$.}
\label{fig:tuning}
\end{figure*}

\paragraph{Effectivness of OT-based Model Fusion.}
To evaluate the effectiveness of OT-based model fusion, we keep Stage 1 unchanged and modify Stage 2 to generate three different versions for comparison. (1) $\textbf{V}_1$: no fusion is conducted in Stage 2; instead, we evaluate the performance of the pruned model from Stage 1; (2) $\textbf{V}_2$: implement vanilla fusion by directly fusing the pruned model with the backdoored model using  equation~\eqref{equ:fusion}; (3) $\textbf{V}_3$ (Ours): the final version, where the full procedures of both Stage 1 and 2 are conducted. Table~\ref{tab:abl_fusion} shows the performances of these three versions on CIFAR-10 across four different attacks. The results validate the effectiveness of aligning neuron weights using OT, where the pruned model inherently achieves a low ASR (or even better) while the high ACC is kept. Although the vanilla fusion ($\textbf{V}_2$) better preserves ACC, it fails in effectively mitigating the backdoor effect.

\begin{figure}[t]
\centering
\includegraphics[width=0.7\columnwidth]{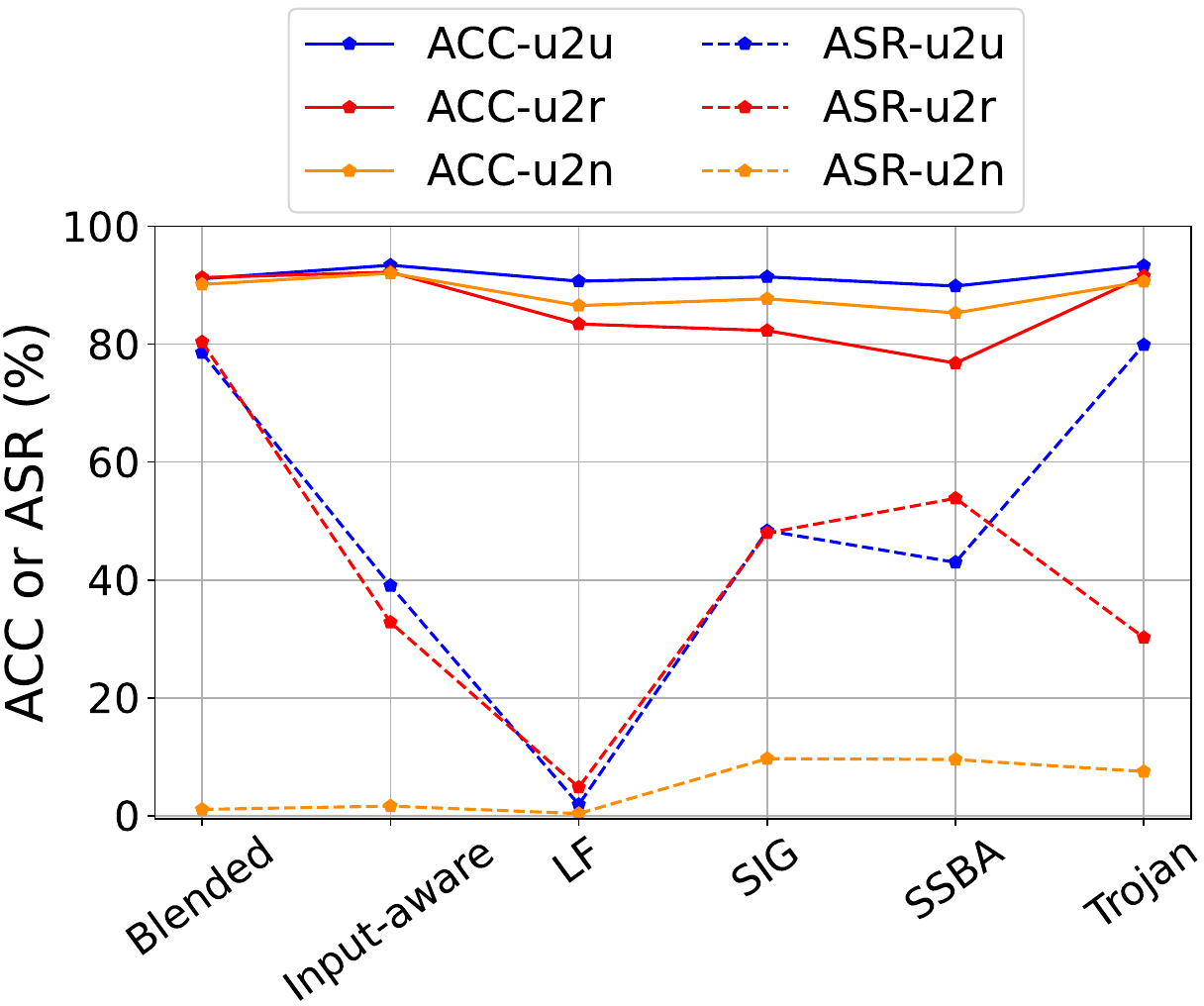} 
\caption{Comparison of different OT distributions. u2u: uniform to uniform transport; u2r: uniform to random transport; u2n(ours): uniform to NWC transport.}
\label{fig:ot_distri}
\end{figure}

\paragraph{Effectiveness of NWC-informed OT.}
We now further verify the important role of NWC-informed OT in achieving optimal defense performance. To do this, we fix the source distribution as a uniform distribution and compare three different initialization schemes for the target distribution in OT. Specifically, we consider three types of distributions: \textit{uniform distribution, random distribution}, and \textit{NWC distribution}.
These schemes are labeled as ``u2u'', ``u2r'', and ''u2n''(Ours), respectively. 
The results are presented in Figure~\ref{fig:ot_distri}. We observe that only ``u2n'', \ie, NWC-informed OT, consistently achieves strong performance across different attacks. In contrast, uniform and random distributions fail to effectively guide neuron weight transport, resulting in poor fusion performance. 

\subsection{Parameter Analysis}
\paragraph{Impact of Different Factors.}
We aim to investigate the impact of various factors on the performance of \nameFramework. These factors include the number of iterative steps $I$, the pruning ratio $\gamma$, the largest label $G$, and the balance coefficient $\lambda$. The experiments are conducted using default settings on CIFAR-10 and BadNets with a 10\% poisoning ratio. The results are shown in \textbf{Figure~\ref{fig:tuning}}. \textbf{Firstly}, in subfigures (a) and (b), we present the results of varying the pruning ratio $\gamma$ from 1\% to 30\% across four numbers of iterative steps $I$ (5, 20, 40, and 80 steps). The two subfigures, showing ACC and ASR respectively, demonstrate that \nameFramework performs well across different settings. A larger pruning ratio tends to require more iterative steps for effective random unlearning. In our setting ($I=20$), $\gamma$ is insensitive in the range of 5\% to 25\%, resulting in successful defense.  
\textbf{Secondly}, in subfigure (c), we show the impact of the balance coefficient $\lambda$ ranging from 0.1 to 1.0. A higher value of $\lambda$ means a more important role the transported model plays in the fusion. We observe that there exists a trade-off between the high ACC from the backdoored model and the low ASR from the transported model, as we assumed before. It suggests setting the $\lambda$ between 0.4 and 0.8 for a successful defense.
\textbf{Lastly}, in subfigure (d), we evaluate the performances with different largest labels $G$ for random data generation to test the impact of class number. Note that 9 is the largest label, in which case the model is trained.
The results reveal that performance remains consistently good across 
$G$ values from 4 to 9, while a smaller class number may fail.  \fin
Overall, our \nameFramework proves to be a robust defense method across various hyperparameter settings.

\section{Conclusion}
\label{sec:conclusion}

In this work, we propose a novel data-free backdoor defense method, \nameFramework, using OT-based model fusion. 
Notably, we provide a new data-free pruning insight by revealing the positive correlation between NWCs when unlearning random noise and poisoned data.
This insight enables us to effectively eliminate the backdoor effect using pruning guided by NWCs in a data-free manner. Then, we propose to combine the high ACC of the backdoored model with the low ASR of the pruned model using the OT-based model fusion. Furthermore, we provide possible explanations for the success of both NWC pruning and OT-based fusion. Extensive experiments across various attacks and datasets confirm the effectiveness of our OTBR method.
A current limitation of this work is its reliance on NWC, which applies only to the post-training scenario.
In future work, we plan to explore the potential of OT-based model fusion for more scenarios, such as in-training defense. 


\bibliography{aaai25}

\appendix
\clearpage
\section{Appendix}

\subsection{Algorithm Outline}
The \nameFramework defense method we propose in this paper can be summarized in Algorithm~\ref{alg:outline}. By inputting a backdoored model $\boldsymbol{\theta}_{bd}$ and numerous hyperparameters, we aim to obtain a repaired clean model in a data-free manner. 
Specifically, we generate a random dataset $\mathcal{D}_{r}$ (line 2) to conduct random unlearning on $\boldsymbol{\theta}_{bd}$ (line 3). Then, we calculate the NWCs (line 4) and use them to obtain a data-free model $\boldsymbol{\theta}_{pn}$ by pruning (line 5). 
After obtaining the pruned model and the NWCs, we conduct Algorithm~\ref{alg:otfusion} with the original backdoored model $\boldsymbol{\theta}_{bd}$ and some hyperparameters $\lambda$ and $p$ (line 7). 
The clean model is obtained from Algorithm~\ref{alg:otfusion}.

\begin{algorithm}[]
    \caption{\nameFramework method}
    \label{alg:outline}
    \raggedright
    {\bf Input}: Backdoored model $\boldsymbol{\theta}_{bd}$, random noise size [A, H, W], batch size $B$, largest class label $G$, the number of iterative steps $I$, pruning ratio $\gamma$, balance coefficient $\lambda$, the first pruned layer $p$\\
    {\bf Output}: Clean model $\boldsymbol{\theta}^*$\\
    \begin{algorithmic} [1]
    \STATE /* \textbf{Stage 1: Random-Unlearning NWC Pruning} */
    \STATE $\mathcal{D}_{r}$ $\leftarrow$ Randomly generate $I \times B$ pairs of noise data.
    \STATE $\boldsymbol{\theta}_{ul}$ $\leftarrow$ Unlearn $\boldsymbol{\theta}_{bd}$ by equation \eqref{equ:unlearn} using $\mathcal{D}_{r}$.
    \STATE Calculate NWCs by equation~\eqref{equ:nwc} using $\boldsymbol{\theta}_{ul}$ and $\boldsymbol{\theta}_{bd}$.
    \STATE $\boldsymbol{\theta}_{pn}$ $\leftarrow$ Prune the largest-$\gamma$ NWC neurons of $\boldsymbol{\theta}_{bd}$. 
    \STATE /* \textbf{Stage 2: Pruned-to-Backdoored OT-based Fusion} */
    \STATE $\boldsymbol{\theta}^{*}$ $\leftarrow$ Conduct Algorithm~\ref{alg:otfusion} with input $(\boldsymbol{\theta}_{pn}, \boldsymbol{\theta}_{bd}, NWCs, \lambda, p)$.
    \end{algorithmic}
\end{algorithm}

\subsection{Complementary Details of Related Work}
\paragraph{Backdoor Defense.}
As illustrated in the \textbf{Related Work} section, backdoor defenses can be categorized into three main types: pre-training, in-training, and post-training defenses. Here, we provide more details on the pre-training and in-training defenses.
\textbf{(1) Pre-training Defense.} This type of defense aims to detect and remove poisoned samples from the training dataset before the model training phase begins. For instance, AC~\cite{chen2018detecting} uses abnormal activation clustering of the target class to identify and filter out poisoned samples, while Confusion Training~\cite{qi2023towards} trains a poisoned model to detect and filter out the poisoned samples that only fit the model's specific characteristics. 
\textbf{(2) In-training Defense.} In this type, defenders focus on the model-training phase, where they have access to both the poisoned training dataset and the training process. ABL~\cite{li2021anti} exploits the observation that loss reduction for poisoned data is faster than that for clean data during the initial training phases, enabling it to isolate and unlearn the poisoned samples; Similarly, D-ST/D-BR~\cite{chen2022effective} identifies a significant sensitivity of poisoned data during the feature representation transformations; DBD~\cite{huang2022backdoor} separates the training of the feature extractor and classifier using different learning techniques to avoid learning the trigger-label correlation.

\subsection{Implementation Details}
\paragraph{Computing Infrastructure.}
All experiments are conducted on a server with 8 \textit{NVIDIA RTX A6000} GPUs and a \textit{Intel(R) Xeon(R) Gold 6226R} CPU. These experiments were successfully executed using less than 49G of memory on a single GPU card. The system version is \textit{Ubuntu 20.04.6 LTS}. We use PyTorch~\cite{paszke2019pytorch} for implementation.

\paragraph{Dataset Details.}
Our experiments are conducted on \textit{CIFAR-10}~\cite{krizhevsky2009learning}, \textit{Tiny ImageNet}~\cite{le2015tiny}, and \textit{CIFAR-100}~\cite{krizhevsky2009learning}. 
We provide a detailed summary of these three datasets in Table~\ref{tab:dataset}. 

\begin{table}[]
\caption{Summary of datasets used in our experiments.}
\label{tab:dataset}
\resizebox{1\linewidth}{!}{
\begin{tabular}{@{}c|cccc@{}}
\toprule
Datasets                                                & \# Input Size         & \# Classes & \begin{tabular}[c]{@{}c@{}}\# Training \\ Images\end{tabular} & \begin{tabular}[c]{@{}c@{}}\# Testing \\ Images\end{tabular} \\ \midrule
CIFAR-10                                                & $3\times 32\times 32$ & 10         & 50,000                                                        & 10,000                                                       \\ \midrule
\begin{tabular}[c]{@{}c@{}}Tiny\\ ImageNet\end{tabular} & $3\times 64\times 64$ & 200        & 100,000                                                       & 10,000                                                       \\ \midrule
CIFAR-100                                               & $3\times 32\times 32$ & 100        & 50,000                                                        & 10,000                                                       \\ \bottomrule
\end{tabular}}
\end{table}

\begin{figure*}[]
    \centering
        \subfigure[BadNets]{\includegraphics[width=0.13\textwidth]{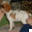}}
        \subfigure[Blended]{\includegraphics[width=0.13\textwidth]{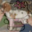}}
        \subfigure[Input-aware]{\includegraphics[width=0.13\textwidth]{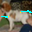}}
        \subfigure[LF]{\includegraphics[width=0.13\textwidth]{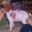}}
        \subfigure[SSBA]{\includegraphics[width=0.13\textwidth]{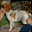}}
        \subfigure[Trojan]{\includegraphics[width=0.13\textwidth]{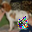}}
        \subfigure[WaNet]{\includegraphics[width=0.13\textwidth]{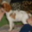}}
    \caption{Examples of all seven backdoor attacks on CIFAR-10 used in our experiments.}
    \label{fig:attack_example}
\end{figure*}

\paragraph{Attack Details.}
We conduct comprehensive evaluations using seven SOTA backdoor attacks, including BadNets~\cite{gu2019badnets}, Blended~\cite{chen2017targeted}, Input-aware~\cite{nguyen2020input}, LF~\cite{zeng2021rethinking}, SSBA~\cite{li2021invisible}, Trojan~\cite{liu2018trojaning} and WaNet~\cite{nguyen2021wanet}. All attacks are conducted using the default settings in BackdoorBench~\cite{wu2022backdoorbench}. 
Figure~\ref{fig:attack_example} shows examples of all seven attacks on CIFAR-10.
Specifically, for BadNets, a 3$\times$3 white square is patched at the bottom-right corner of the images for CIFAR-10 and CIFAR-100, while a 6$\times$6 white square is used for Tiny ImageNet. For Blended, a Hello-Ketty image is blended into the images with a transparent ratio of 0.2. 
We choose a 10\% poisoning ratio and the 0$^{th}$ label as the default settings to conduct attacks and test all defenses, following previous works~\cite{zhu2023enhancing,wei2023shared}. Other poisoning ratios of 5\%, 1\%, 0.5\%, and 0.1\% are used only for testing our proposed method.

\paragraph{Defense Details.}
We compare our proposed \nameFramework method with six SOTA defense methods, including Fine-pruning (FP)~\cite{liu2018fine}, NAD~\cite{li2021neural}, ANP~\cite{wu2021adversarial}, i-BAU~\cite{zeng2021adversarial}, RNP~\cite{li2023reconstructive}, and CLP~\cite{zheng2022data}. Note that only CLP can be conducted in a data-free manner similar to \nameFramework, while the other five defenses are all data-dependent. Therefore, we follow the common setting in the post-training scenario, providing 5\% clean data for those methods. The learning rate for all methods is set to $10^{-2}$, and the batch size is set to 256. For RNP, the clean data ratio is adjusted to 0.5\% since we found that RNP did not perform well with the 5\% setting. For \nameFramework, we set the number of iterative steps $I$ to 20 during random unlearning. The generated noise conditions are similar to CIFAR-10, with image shapes set to [3, 32, 32] and the largest label $G$ set to the output dimension of the backdoored model. The default pruning ratio $\gamma$ is set to 5\%, and the balance coefficient $\lambda$ is set to 0.5. 

\subsection{More Experiments}
\paragraph{Computational Cost.}
We now analyze the computational cost of our proposed \nameFramework in terms of runtime on CIFAR-10 and Tiny ImageNet. The operation time for each defense step of  \nameFramework and CLP is illustrated in Table~\ref{tab:compute_time_step}. We observe that for \nameFramework, both Stage 1 (Random Unlearning + Pruning) and Stage 2 (OT + Fusion) are highly efficient, with runtimes of less than 2 seconds. However, CLP, as the compared data-free defense, is much slower than \nameFramework in calculating the UCLC (\textit{Upper bound of the Channel Lipschitz Constant}) and in pruning, \eg, $8.04s > (1.96s + 1.16s)$ in CIFAR-10. The reason is that CLP needs to calculate each neuron's UCLC individually, while our \nameFramework can calculate NWC and conduct OT-based Fusion layer by layer.

\begin{table}[]
\caption{Operation time of each step on \nameFramework and CLP.}
\label{tab:compute_time_step}
\resizebox{1\linewidth}{!}{
\begin{tabular}{@{}cccc@{}}
\toprule
Defense               & Defense Step                                                          & CIFAR-10 & Tiny ImageNet \\ \midrule
\multirow{2}{*}{OTBR} & \begin{tabular}[c]{@{}c@{}}Random Unlearning\\ + Pruning\end{tabular} & 1.96s    & 0.91s         \\ \cmidrule(l){2-4} 
                      & OT + Fusion                                                           & 1.16s    & 1.29s         \\ \midrule
CLP                   & \begin{tabular}[c]{@{}c@{}}Calculate UCLC\\ + Pruning\end{tabular}    & 8.04s    & 8.09s         \\ \bottomrule
\end{tabular}}
\end{table}

\paragraph{Performance Under Different Poisoning Ratios.}
To evaluate the effectiveness of \nameFramework under different poisoning ratios, we test the BadNets attack on both CIFAR-10 and Tiny ImageNet datasets. The testing range includes poisoning ratios of 10\%, 5\%, 1\%, 0.5\%, and 0.1\%. It is important to note that a smaller poisoning ratio results in a more hidden attack. 
As shown in Table~\ref{tab:poison_ratio}, \nameFramework consistently succeeds in defense regardless of the poisoning ratio.

\begin{table*}[]
\caption{Performance under different poisoning ratios on BadNet (\%).}
\label{tab:poison_ratio}
\centering
\resizebox{\linewidth}{!}{
\begin{tabular}{cc|cc|cc|cc|cc|cc}
\hline
\multicolumn{2}{c|}{Poisoning Ratio}                                                                    & \multicolumn{2}{c|}{10\%} & \multicolumn{2}{c|}{5\%} & \multicolumn{2}{c|}{1\%} & \multicolumn{2}{c|}{0.5\%} & \multicolumn{2}{c}{0.1\%} \\ \hline
\multicolumn{1}{c|}{Datasets}                                                                  & Metric & No Defense     & Ours     & No Defense    & Ours     & No Defense    & Ours     & No Defense     & Ours      & No Defense     & Ours     \\ \hline
\multicolumn{1}{c|}{\multirow{2}{*}{CIFAR-10}}                                                 & ACC    & 91.32          & \greencell{90.11}    & 92.64         & \greencell{89.25}    & 93.14         & \greencell{89.49}    & 93.76          & \greencell{91.04}     & 93.61          & \greencell{90.83}    \\
\multicolumn{1}{c|}{}                                                                          & ASR    & 95.03          & \greencell{1.08}     & 88.74         & \greencell{0.28}     & 74.73         & \greencell{0.17}     & 50.06          & \greencell{0.18}      & 1.23           & \greencell{1.21}     \\ \hline
\multicolumn{1}{c|}{\multirow{2}{*}{\begin{tabular}[c]{@{}c@{}}Tiny \\ ImageNet\end{tabular}}} & ACC    & 56.23          & \greencell{54.13}    & 56.22         & \greencell{52.90}    & 57.15         & \greencell{48.29}    & 57.40          & \greencell{54.98}     & 57.51          & \greencell{52.31}    \\
\multicolumn{1}{c|}{}                                                                          & ASR    & 100.00         & \greencell{0.00}     & 99.85         & \greencell{0.00}     & 95.01         & \greencell{0.01}     & 91.26          & \greencell{1.98}      & 33.52          & \greencell{2.24}     \\ \hline
\end{tabular}}
\end{table*}

\paragraph{Performance on Different Model Structures.}
We evaluate the effectiveness of \nameFramework on different model structures except for the default PreAct-ResNet18. Specifically, we consider the VGG19-BN~\cite{simonyan2014very} and ResNet18~\cite{he2016deep} models, comparing  \nameFramework with CLP on CIFAR-10. The results, shown in Table~\ref{tab:perform_model}, reveal that \nameFramework successfully defends against various attacks on both models, while CLP fails to defend against Trojan in VGG19-BN. This validates the effectiveness of \nameFramework across different model structures.

\begin{table*}[]
\caption{Performances of different models on CIFAR-10.}
\label{tab:perform_model}
\centering
\begin{tabular}{@{}c|c|cc|cc|cc@{}}
\toprule
\multirow{2}{*}{Model}    & \multirow{2}{*}{Attacks} & \multicolumn{2}{c|}{No Defense} & \multicolumn{2}{c|}{CLP} & \multicolumn{2}{c}{\textbf{\nameFramework}} \\ \cmidrule(l){3-8} 
                          &                          & ACC            & ASR            & ACC         & ASR        & ACC         & ASR        \\ \midrule
\multirow{2}{*}{VGG19-BN} & LF                       & 83.27          & 93.83          & \greencell{81.99}       & \greencell{14.61}      & \greencell{80.73}       & \greencell{18.37}      \\ \cmidrule(l){2-8} 
                          & Trojan                   & 91.57          & 100.00         & 90.56       & 99.70      & \greencell{88.06}       & \greencell{0.11}       \\ \midrule
\multirow{2}{*}{ResNet18} & BadNets                  & 91.72          & 94.80          & \greencell{91.49}       & \greencell{9.02}       & \greencell{84.64}       & \greencell{8.07}       \\ \cmidrule(l){2-8} 
                          & Input-aware              & 91.75          & 92.96          & \greencell{91.28}       & \greencell{1.66}       & \greencell{91.06}       & \greencell{2.20}       \\ \bottomrule
\end{tabular}
\end{table*}

\subsection{Further Analysis}

\begin{figure*}[t]
\centering
\includegraphics[width=0.38\linewidth]{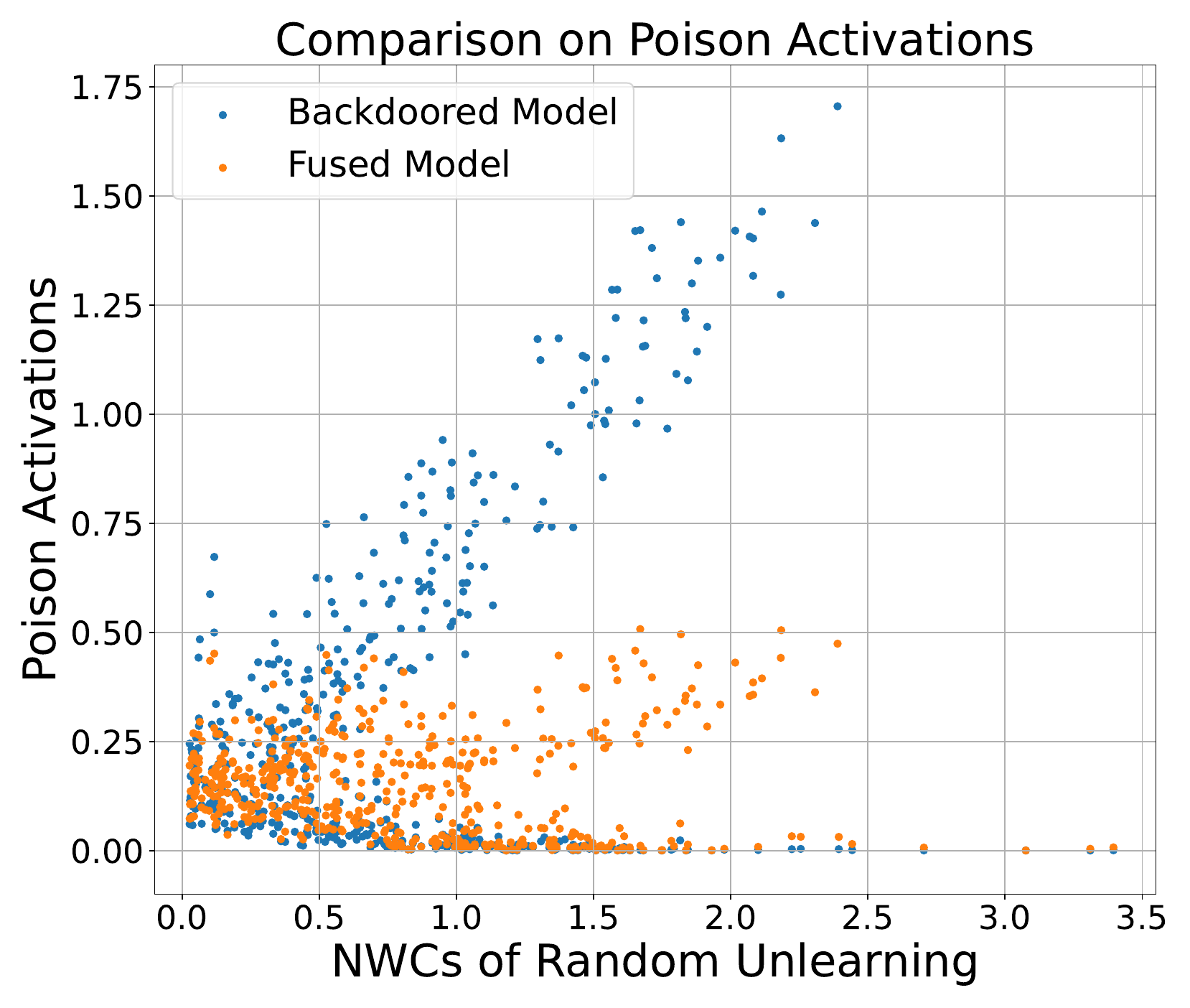}
\includegraphics[width=0.38\linewidth]{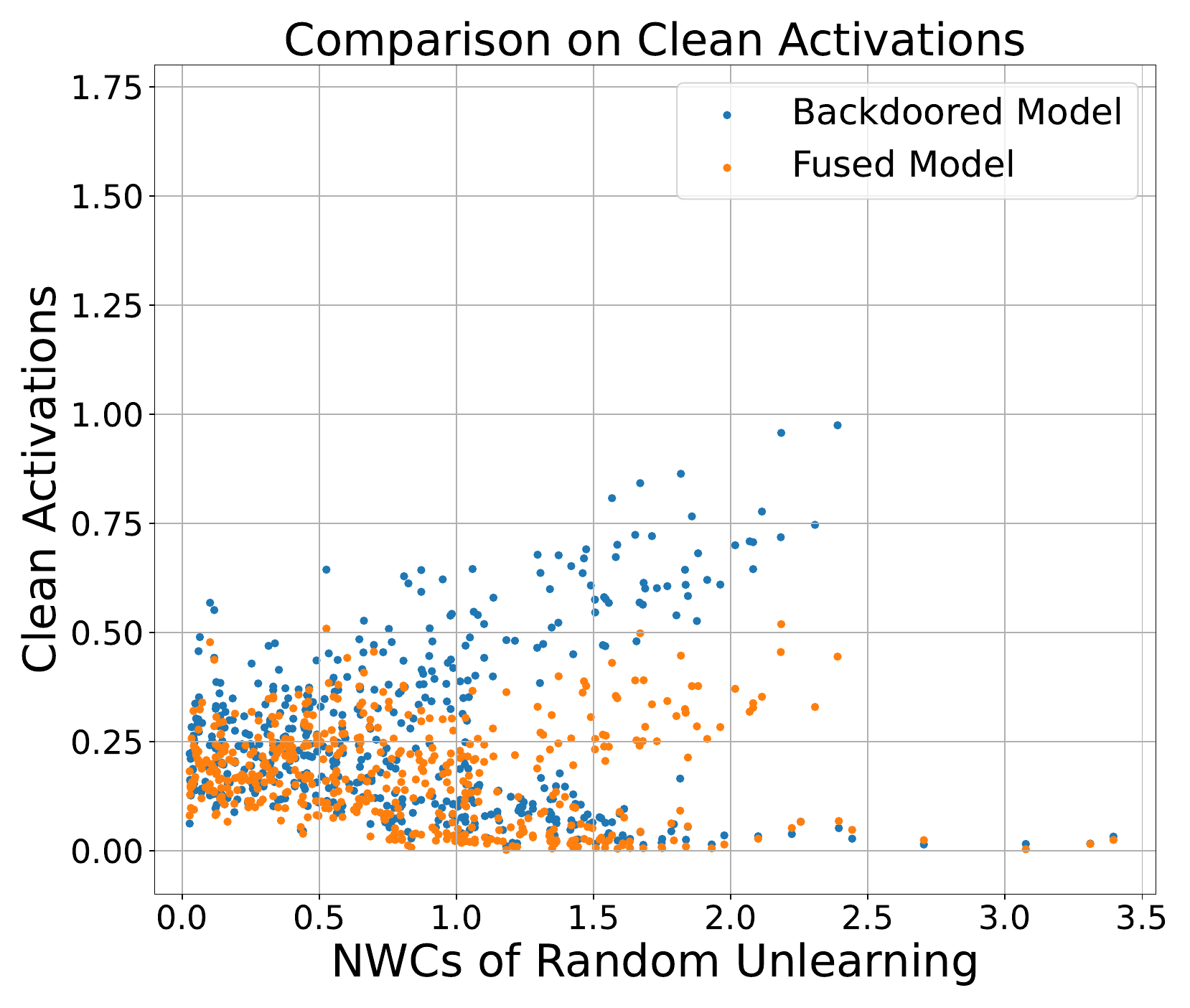}
\caption{Illustration of neuron-level activations on backdoored and fused models. \textbf{Left}: the activations of backdoored and fused models with poisoned input; \textbf{Right}: the activations of backdoored and fused models with clean input. The last convolutional layer with \textit{relu} activation function is used for illustration.}
\label{fig:activation}
\end{figure*}

\begin{figure*}[]
\centering
\includegraphics[width=0.4\linewidth]{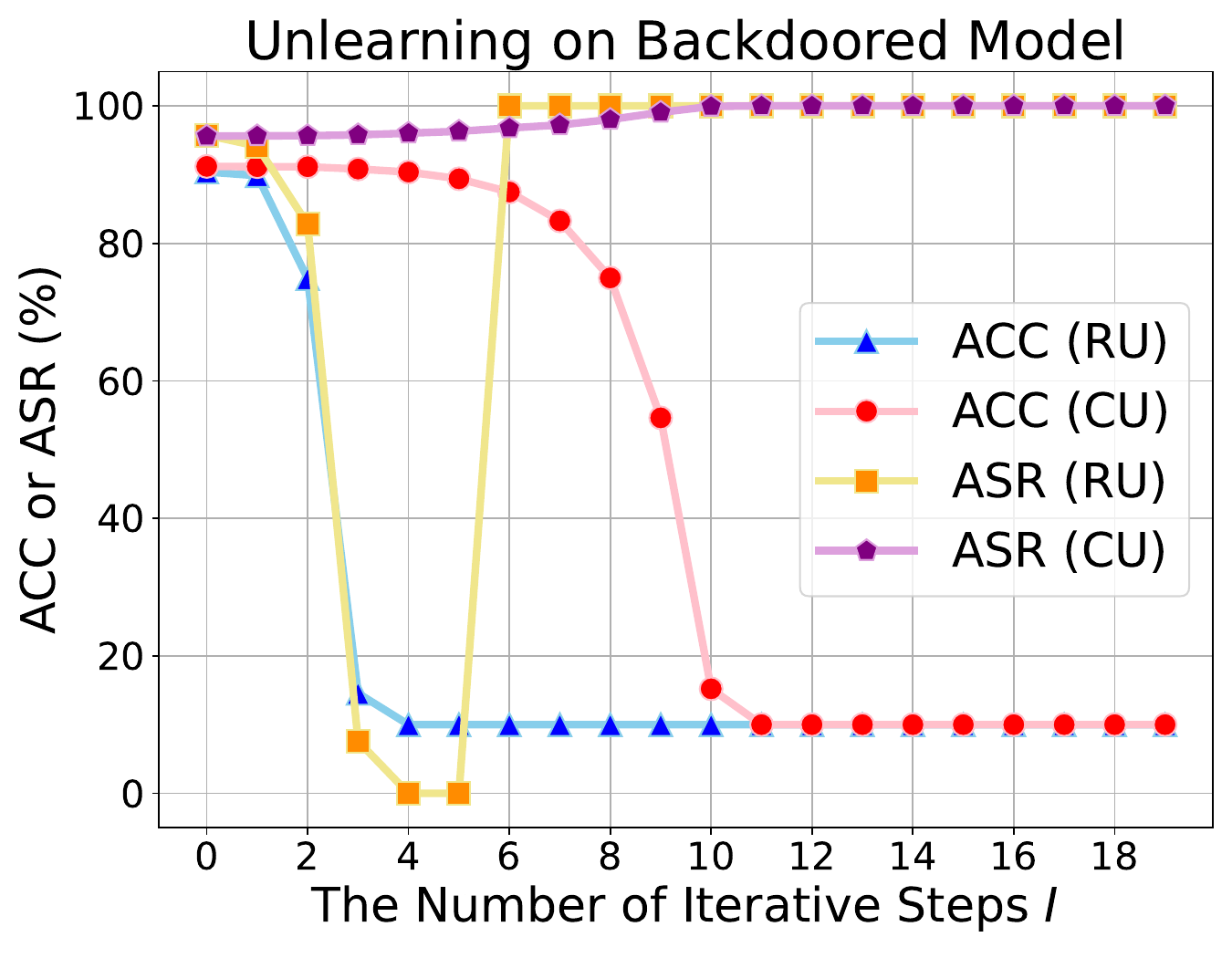} 
\caption{Comparisons of the unlearning process on clean input and random input on the backdoored model. RU: \textit{Random Unlearning}; CU: \textit{Clean Unlearning}.}
\label{fig:unlearning}
\end{figure*}

\paragraph{Why Can Random-Unlearning NWCs Expose Backdoors? }
To answer this question, we take a close look at the unlearning process using both random data and clean data as inputs.  Figure~\ref{fig:unlearning} compares the changes in ACC and ASR between clean unlearning and random unlearning. We observe that during the first few iterative steps, unlearning random noise adversely affects both ACC and ASR, while unlearning clean data only declines ACC, which is intuitively reasonable. However, after more unlearning steps, the performance for both types of unlearning converges, resulting in high ASR and low ACC. From the above derivation, we deduce that both types of unlearning optimization achieve similar results after a certain number of iterative steps, which suggests similar weight-change behavior in neurons. Therefore, by replacing clean unlearning with random unlearning, we can also expose the backdoor-related neurons using NWCs, as demonstrated in \cite{lin2024unveiling}.

\paragraph{Observations in Activations.}
The activation comparisons between the backdoored model and the final fused model are shown in Figure~\ref{fig:activation}.
The left subfigure represents poisoned inputs, while the right subfigure represents clean inputs. We observe that neurons with high-NWCs, which are more backdoor-related, experience a similar decrease in activation levels for both poisoned and clean inputs, while a significant drop occurs for poisoned inputs. This indicates that the defense is successful while having a minimal impact on clean functionality. 

\end{document}